\documentclass[aip,notitlepage, floatfix, jcp, preprint]{revtex4-1}
\usepackage{amssymb}
\usepackage{amsfonts}
\usepackage{amsmath}
\usepackage{graphicx}
\usepackage{enumerate}
\usepackage[usenames,dvipsnames]{color}
\usepackage{ulem}

\input{tcilatex}
\begin{document}

\title{A dynamical theory of nucleation for colloids and macromolecules}
\author{James F. Lutsko}
\affiliation{Center for Nonlinear Phenomena and Complex Systems, Code Postal 231,
Universit\'{e} Libre de Bruxelles, Blvd. du Triomphe, 1050 Brussels, Belgium}
\email{jlutsko@ulb.ac.be}
\homepage{http://www.lutsko.com}
%\pacs{64.60.Q-, 82.60.Nh, 05.40.-a}

\begin{abstract}
A general theory of nucleation for colloids and macromolecules in solution
is formulated within the context of fluctuating hydrodynamics. A formalism
for the determination of nucleation pathways is developed and stochastic
differential equations for the evolution of order parameters are given. The
conditions under which the elements of Classical Nucleation Theory are
recovered are determined. The theory provides a justification and extension
of more heuristic equilibrium approaches based solely on the free energy. It
is illustrated by application to the low-concentration/high-concentration
transition in globular proteins where a novel two-step mechanism is
identified where the first step involves the formation of long-wavelength
density fluctuations and the second step is the actual nucleation event
occurring within the fluctuation.
\end{abstract}

\date{\today }
\maketitle

\section{Introduction}

Nucleation is a fundamental physical process of importance in fields as
diverse as chemistry, materials science, biology and cosmology. Our basic
understanding of it goes back to Gibbs' discussion of the nucleation of a
liquid droplet in a metastable vapor, see e.g. discussion in Ref. \onlinecite{Kashchiev}. The physics is
governed by the fact that the excess free energy of a cluster of new phase
relative to the mother phase has a negative contribution that scales as the
volume and a positive contribution that scales as the surface area. The
former is due to the fact that the bulk new phase has lower free energy than
the bulk mother phase while the latter arises from the free energy cost of
the interfacial region. When the cluster is small, the surface term
dominates and the excess free energy of the cluster is positive making the
cluster thermodynamically disfavored. When the cluster is large, the volume
term dominates, the cluster has lower free energy than the vapor and so
cluster growth is favored. Separating the two regimes is the metastable
critical cluster. In Classical Nucleation Theory (CNT) it is assumed that
the cluster is spherical, that its interior is in the bulk new-phase state
and that the surface tension is the same as for a planar interface of the
coexisting new and mother phases so that the free energy of the cluster as a
function of its radius can be calculated giving a quantitative picture of
homogeneous nucleation\cite{Kashchiev}.

The dynamics of nucleation - the nucleation rate, the determination of the
nucleation pathway and the nature of the post-critical growth of clusters -
is a more involved subject. In CNT, the starting point for understanding
nucleation rates is the monomer attachment/detachment picture that leads to
the Becker-D\"{o}ring equations\cite{BeckerDoring,Kashchiev}. These are
simply rate equations for number of clusters of size $N$ based on the
processes of monomer attachment and detachment. Following Frenkel,
Zeldovich, and others\cite{Kashchiev}, these equations can be converted into
a Fokker-Planck equation for the evolution of the distribution of clusters.
All of these formalisms can be used to estimate nucleation rates. The
physics enters via the monomer attachment and monomer detachment rates which
in general depend on the thermodynamic driving force (the derivative of the
CNT free energy with respect to cluster size) and the dynamics of the
surrounding mother phase. In many cases, adequate approximations are
available that express the nucleation rate in terms of $e^{-\beta \Delta
\Omega _{c}}$ where $\beta =1/k_{B}T$, $k_{B}$ is Boltzmann's constant, $T$
is the temperature and $\Delta \Omega _{c}$ is the excess free energy of the
critical cluster.

Although physically appealing, this formalism suffers from a number of
deficiencies. First and foremost is the well-known fact that the nucleation
rates estimated from CNT are often many of orders of magnitude too large or
too small\cite{Granasy} (see also Ref. \onlinecite{Kashchiev}, Fig. 13.8).
This is largely attributed to a poor estimate of the free energy barrier:
more sophisticated methods allow for better estimates of the barriers. A common way to get improved estimates of the
barriers is via classical Density Functional Theory (DFT)\cite%
{Evans79,HansenMcdonald,LutskoAdvChemPhysDFT} (for early examples of this
approach see Refs. \onlinecite{OxtobyEvans,talanquer:5190}). However, the
use of DFT raises new conceptual questions. The monomer
attachment/detachment picture requires estimates of the rates for all
cluster sizes which in turn means estimating "free energies" for all
clusters. But, DFT is an equilibrium theory that can only describe - at best
- stationary states on the free energy surface. This fact follows trivially
since the procedure to determine free energies via DFT begins with the
extremization of the free energy to get the structure of the cluster and
naturally must result in a minimum, maximum or saddle point but it is also
connected to the theorems that underlay DFT, which only attach significance
to minima of the free energy functional\cite{MerminDFT,HansenMcdonald}. This
therefore raises the question as to whether DFT can be used to determine the
nucleation pathway - i.e. the sequence of necessarily nonequilibrium states
leading from the pure unstable phase to the critical cluster.

Putting aside the conceptual questions, there are currently two primary
approaches to determining nucleation pathways using DFT. The first, and
oldest, is to extremize the free energy under some sort of constraint that
stabilizes non-critical clusters:\ in effect, imposing a stabilizing
external field\cite{OxtobyEvans,talanquer:5190,LutskoBubble2}. The problem
with this approach is that there are numerous possible ways to implement it.
To take a simple example, one could minimize the free energy at fixed
 radius so that the pathway would be mapped out by varying the
radius which acts as the reaction coordinate (or "order parameter").
However, one could equally well minimize at fixed number of molecules in the
cluster, $N$, so that this becomes the order parameter. These do not
necessarily give the same result since, e.g., the number of molecules can
increase with the  radius actually decreasing provided the
interfacial width or interior density of the cluster increases enough to
compensate; indeed, this actually occurs in some calculations\cite%
{Lutsko2011a}. Notwithstanding this problem, the constraint approach
continues to be widely used\cite{Ghosh, Corti_PRL, Corti1, *Corti2,
EvansArcherNucleation}. A more recent alternative is to search for steepest
descent paths on the free energy surface\cite%
{LutskoBubble1,LutskoBubble2,MFEP_App,Iwa,Philippe}. However, here there is
also an ambiguity as steepest descent requires a measure of distance in
parameter space and there is \textit{a priori} no unique prescription for this
measure\cite{LutskoBubble1}.

It is reasonable to suppose that all of the ambiguity underlying the DFT
approaches is due to the fact that DFT is an equilibrium theory whereas
nucleation is fundamentally a nonequilibrium, fluctuation-driven process.
Viewed in this way, one can question other aspects of CNT as well, such as
whether it is really necessary that the system even passes through the
critical cluster since the free energy may not be the only factor
controlling the nonequilibrium evolution. Similarly, one can also ask why
the pre-critical process does not seem to depend on fluid-kinetics while
these are crucial in driving post-critical growth. (This question has also been addressed recently by Peters\cite{Peters} using CNT free energy models). All of this suggests
that one would like a unified description with the following properties:\ it
is inherently nonequilibrium in nature, it is dynamical; it allows for the
determination of the nucleation pathway from the dynamics, it allows for the
determination of nucleation rates and it gives a description of
post-critical growth. Of course, one would expect that it involves free energies
 in a natural way without assuming equilibrium and that under some limit
the CNT would result thus also helping to clarify its applicability.
Additionally, one would hope that such a formalism would be sufficiently
flexible to be applied to more complex problems involving multiple order
parameters and to determine the relative probability of completely different
nucleation pathways (e.g. whether crystals form directly from solution or
whether they form in two steps, passing first through a dense amorphous phase%
\cite{LutskoAdvChemPhys}).

The aim of this paper is to illustrate an approach that can satisfy these
requirements. Attention will primarily be focused on nucleation from a vapor
or liquid phase, but the basic ideas could be extended to other systems in
obvious ways. The development is based on  Brownian Dynamics
where molecules move according to Newton's laws while being subject to a
frictional force as well as fluctuating forces. This is a simple model for
colloids and the important case of macromolecules in solution in which cases
the friction and the fluctuations come from the bath/solvent and has the
virtue of allowing for useful simplifications. As described in Section II,
the fundamental level of description adopted for the description of
nucleation is fluctuating hydrodynamics, in which the free energy enters the
dynamics via a local equilibrium expression for the pressure, resulting in a
set of stochastic differential equations (SDE's). The Most Likely Path (MLP)
is introduced as a means of characterizing the nucleation pathway via a 
\emph{deterministic} equation derived from the stochastic model. In Section
III a reduced description in terms of order parameters is developed. It is
shown that in the weak-noise limit the MLP is simply gradient descent on the
free energy surface with a metric uniquely determined by the dynamics and
that it necessarily passes through the critical cluster thus making contact
with the CNT picture. Finally, in Section IV three applications are
discussed:\ nucleation of high concentration droplets from a low
concentration solution of globular proteins; the reverse transition of high
concentration solution to low concentration bubbles in an under-saturated
solution and finally the specialization to a single order parameter so as to
recover well-known results from CNT. For the first application - droplet nucleation - calculations lead to an unexpected but physically appealing result wherein it is found that the coupling of fluctuations, fluid dynamics and thermodynamics implies that nucleation is most likely to occur within regions of enhanced density due to long-wavelength density fluctuations. The final Section summarizes the results and discusses
possibilities for further developments. A brief summary of some of these
results has appeared previously\cite{Lutsko_JCP_2011_Com}.

\section{The model}

\subsection{Dynamics}

The system consists of a collection of particles with positions and momenta $%
\mathbf{q}_{i}$,$\mathbf{p}_{i}$ interacting via a potential $V$.
Additionally, the particles interact with a bath/solvent of small particles
and this is described via a frictional drag and a fluctuating force, both characterized by the constant $\gamma$,%
\begin{equation}
\overset{\cdot }{\mathbf{q}}_{i}=\mathbf{p}_{i}/m,\;\;\overset{\cdot }{%
\mathbf{p}}_{i}=-\frac{\partial V}{\partial \mathbf{q}_{i}}-\gamma \mathbf{p}%
_{i}+\varepsilon \sqrt{2\gamma mk_{B}T}\mathbf{f}_{i}\left( t\right)
\end{equation}%
where all the components of the force are independent,%
\begin{equation}
\left\langle \mathbf{f}_{i}\left( t\right) \mathbf{f}_{j}\left( t^{\prime
}\right) \right\rangle =\mathbf{1}\delta _{ij}\delta \left( t-t^{\prime
}\right)
\end{equation}%
The coefficient of the noise term is determined by demanding that the system
admits of an equilibrium Boltzmann distribution thus forcing a
fluctuation-dissipation relation. Here and in the following, an additional
parameter, $\varepsilon $, is included to allow for the discussion of a
weak-noise limit. The physicality of such a limit will be discussed below.

Defining the local density and momentum density respectively as%
\begin{align}
\widehat{\rho }\left( \mathbf{r};t\right) & =\sum_{i}\delta \left( \mathbf{r}%
-\mathbf{q}_{i}\right) \\
\widehat{\mathbf{j}}\left( \mathbf{r};t\right) & =\sum_{i}\mathbf{p}%
_{i}\delta \left( \mathbf{r}-\mathbf{q}_{i}\right)  \notag
\end{align}%
one sees that these satisfy the exact equations%
\begin{align}
\frac{\partial \widehat{\rho }\left( \mathbf{r};t\right) }{\partial t}& =-%
\frac{1}{m}\mathbf{\nabla \cdot }\widehat{\mathbf{j}}\left( \mathbf{r}%
;t\right)  \label{prim} \\
\frac{\partial \widehat{\mathbf{j}}\left( \mathbf{r};t\right) }{\partial t}&
=\mathbf{\nabla \cdot }\sum_{i}\frac{\mathbf{p}_{i}\mathbf{p}_{i}}{m}\delta
\left( \mathbf{r}-\mathbf{q}_{i}\right) -\sum_{i}\frac{\partial V}{\partial 
\mathbf{q}_{i}}\delta \left( \mathbf{r}-\mathbf{q}_{i}\right)  \notag \\
& +\gamma \widehat{\mathbf{j}}\left( \mathbf{r};t\right) +\varepsilon \sqrt{%
2\gamma mk_{B}T\widehat{\rho }\left( \mathbf{r};t\right) }\mathbf{F}\left( 
\mathbf{r;}t\right)  \notag
\end{align}%
with%
\begin{equation}
\left\langle \mathbf{F}\left( \mathbf{r;}t\right) \mathbf{F}\left( \mathbf{r}%
^{\prime }\mathbf{;}t^{\prime }\right) \right\rangle =\delta \left(
t-t^{\prime }\right) \delta \left( \mathbf{r}-\mathbf{r}^{\prime }\right) 
\mathbf{1}
\end{equation}%
Note that the actual form of the noise term that occurs in the momentum
equation is 
\begin{equation}
\varepsilon \sqrt{2\gamma mk_{B}T}\sum_{i}\delta \left( \mathbf{r}-\mathbf{q}%
_{i}\right) \mathbf{f}_{i}\left( t\right)
\end{equation}%
However, it is well-known that a noise term of the form $B_{ij}\xi _{j}(t)$ (with summation over repeated indices)
can be replaced by another of the form $C_{ij}\xi _{j}(t)$ without changing
the statistical properties of the dependent variables provided that
coefficients are related by $B_{il}B_{jl}=C_{il}C_{jl}$, i.e. the
autocorrelation of the noise is unchanged\cite{Gardiner,Dean}.

There are two sources of stochasticity in this problem. The obvious source
is the fluctuating force that represents the bath or solvent. The second
arises because one is not interested in following a single trajectory of the
system but, as usual, want to consider an average over an ensemble of
systems. In equilibrium, the ensemble is drawn from a known distribution
(the canonical distribution). The distribution for systems out of
equilibrium is generally time-dependent and one must specify the ensemble for a particular point in time, e.g. a distribution of
initial conditions. In the present case, because of the two sources of
noise, one must carry out this procedure in two steps: first, the
set of values of the fluctuating forces acting on the colloids at each instant in
time is specified. Then, this fixed set of values for the noise simply
acts as an external field coupled to the colloid dynamics  and nonequilibrium
statistical mechanics can be developed in the usual way as an average over
an ensemble of initial conditions of the particles. Then, having developed
whatever level of description is of interest, a second average is performed
over the noise. As a result, one can go from the exact balance equations,
Eq.(\ref{prim}), to a description of ensemble-averaged quantities.

To fix the notation, let $\Gamma(t)=\{\mathbf{q}_i,\mathbf{p}_i\}_{i=1}^N$
represent the collection of positions and momenta of all $N$ large particles
at time $t$ and to be more explicit one can write $\Gamma(t;\Gamma(0))$
since the phase variables at time $t$ are completely determined by their
values at some earlier time. Similarly, one can write for the density, for
example, $\widehat{\rho }\left( \mathbf{r};t\right) \equiv \widehat{\rho }\left( 
\mathbf{r};\Gamma(t;\Gamma(0))\right)$. The probability for the phase
variables to have particular initial values is denoted $f\left( \Gamma(0)
\right)$ and the ensemble-averaged density is then 
\begin{eqnarray}
\rho(\mathbf{r},t) &=& \int \widehat{\rho }\left( \mathbf{r}%
;\Gamma(t;\Gamma(0))\right) f\left( \Gamma(0) \right)d\Gamma(0) \\
&=& \int \widehat{\rho }\left( \mathbf{r};\Gamma\right) f\left( \Gamma;t
\right)d\Gamma  \notag
\end{eqnarray}
where in the second line the time-dependence has been moved to the
distribution function\cite{McLennan}. Notice that the ensemble-averaged
density and momentum density are all conditional on the fluctuating forces
which for now just act as external forces. At this point, the usual methods
of nonequilibrium statistical mechanics can be used to develop a
hydrodynamic description - either a mesoscopic fluctuating hydrodynamics or
a macroscopic, deterministic hydrodynamics, the former having the form 
\begin{eqnarray}
&\frac{\partial \rho \left( \mathbf{r}\right) }{\partial t}+\frac{1}{m}%
\mathbf{\nabla \cdot j}\left( \mathbf{r}\right) =0 \\
&\frac{\partial \mathbf{j}\left( \mathbf{r}\right) }{\partial t}+\mathbf{%
\nabla \cdot j}\left( \mathbf{r}\right) \mathbf{j}\left( \mathbf{r}\right)
/\rho \left( \mathbf{r}\right) +\rho \mathbf{\nabla}p(\mathbf{r},t)+\mathbf{%
\nabla \cdot \Pi }\left( \mathbf{r}\right) =-\gamma \mathbf{j}\left( \mathbf{%
r}\right) +\varepsilon \sqrt{2m\gamma k_{B}T\rho \left( \mathbf{r}\right) } \mathbf{\xi }%
\left( \mathbf{r};t\right)  \notag
\end{eqnarray}%
where $\rho \left( \mathbf{r}\right) $ and $\mathbf{j}\left( \mathbf{r}%
\right) $ are the ensemble-averaged local density and momentum density and $%
\Pi(\mathbf{r},t)$ is the dissipative stress tensor which for the case of
fluctuating hydrodynamics will have both a deterministic part  and
a fluctuating part. The other new term that appears here is $p(\mathbf{r},t)$
which is the local pressure. Note that in the present case, the bath acts
 as a thermostat and so temperature fluctuations and heat
transport are neglected, although in principle the model could be extended
to include them.

The final approximation involves the pressure and can be motivated in
different ways but always amounts to an assumption of local equilibrium. At
its most heuristic, one starts with the Gibbs-Duhem equation for a
single-component system,$Nd\mu = -SdT+Vdp$. For an isothermal system, this
gives $dp=\rho d\mu$ where $\rho$ is the density. Assuming local-equilibrium
this is generalized to 
\begin{equation}
dp(\mathbf{r},t)=\rho(\mathbf{r},t)d\mu(\mathbf{r},t)=\rho(\mathbf{r},t)d%
\frac{\delta F\left[ \rho \right] }{\delta \rho \left( \mathbf{r}\right) }
\end{equation}
which is a widely used approximation(see, e.g., Ref. \onlinecite{Kashchiev},
Section 8.4). Substituting into the equations for the density and momentum
gives 
\begin{eqnarray}
&\frac{\partial \rho \left( \mathbf{r}\right) }{\partial t}+\frac{1}{m}%
\mathbf{\nabla \cdot j}\left( \mathbf{r}\right) =0 \\
&\frac{\partial \mathbf{j}\left( \mathbf{r}\right) }{\partial t}+\mathbf{%
\nabla \cdot j}\left( \mathbf{r}\right) \mathbf{j}\left( \mathbf{r}\right)
/\rho \left( \mathbf{r}\right) +\rho \left( \mathbf{r}\right) \mathbf{\nabla 
}\frac{\delta F\left[ \rho \right] }{\delta \rho \left( \mathbf{r}\right) }+%
\mathbf{\nabla \cdot \Pi }\left( \mathbf{r}\right) =-\gamma \mathbf{j}\left( 
\mathbf{r}\right) +\varepsilon \sqrt{2m\gamma k_{B}T\rho \left( \mathbf{r}\right) } 
\mathbf{\xi }\left( \mathbf{r};t\right)  \notag
\end{eqnarray}%
An alternative way to motivate this is to evaluate the pressure under the assumption of a local equilibrium (i.e. Maxwellian)  velocity distribution 
and using the local-equilibrium version of the
first YBG equation\cite{EvansArcher} to rearrange the resulting virial expression for the pressure. The
same system of equations has been developed starting with the Boltzmann
equation\cite{ArcherJCP}. Chavanis\cite{Chavanis_Deriv} has discussed the
derivation of these equations for coarse-grained quantities (i.e. averaged
over small volumes and time intervals) as well as their derivation based on
the general Landau-Lifshitz fluctuation formalism\cite{Chavanis_Deriv}. Chavanis also
discusses the difference between this model and the Cahn-Hillard dynamics,
the main point being that the latter does not include the density in the
noise amplitude and the free-energy term. However it is found, the most
important assumption is clearly the introduction of the local equilibrium
expression for the pressure since this is what brings the free energy into
the picture. This assumption of local equilibrium is, on the one hand,
heuristic and represents an uncontrolled approximation. On the other hand,
it is widely used in nonequilibrium statistical mechanics (e.g. a similar
assumption underlies the generalized Enskog equation\cite{McLennan}) and
similar approximations are even used for dissipative granular systems. Thus,
while it cannot be rigorously justified it does represent the ``state of the
art''.

Having developed the general framework, it is now specialized to the case of
strong dissipation due to the bath. This assumption greatly simplifies the
analysis as it means that (a) the dissipative stresses can be neglected
since they will act more slowly than the friction due to the bath and (b)
the momentum current will always be small so that the quadratic convective
term can be neglected. The result is 
\begin{align}
\frac{\partial \rho \left( \mathbf{r}\right) }{\partial t}+\frac{1}{m}%
\mathbf{\nabla \cdot j}\left( \mathbf{r}\right) & =0 \\
\frac{\partial \mathbf{j}\left( \mathbf{r}\right) }{\partial t}+\rho \left( 
\mathbf{r}\right) \mathbf{\nabla }\frac{\delta F\left[ \rho \right] }{\delta
\rho \left( \mathbf{r}\right) }& =-\gamma \mathbf{j}\left( \mathbf{r}\right)
+\varepsilon \sqrt{2\gamma mk_{B}T\rho \left( \mathbf{r}\right) }\mathbf{\xi }\left( 
\mathbf{r};t\right)  \notag
\end{align}%
The linearity allows us to eliminate the momentum giving%
\begin{equation}
\frac{1}{\gamma }\frac{\partial ^{2}\rho \left( \mathbf{r}\right) }{\partial
t^{2}}+\frac{\partial \rho \left( \mathbf{r}\right) }{\partial t}=\frac{%
k_{B}T}{\gamma m}\mathbf{\nabla \cdot }\left( \rho \left( \mathbf{r}\right) \mathbf{%
\nabla }\frac{\delta \beta F\left[ \rho \right] }{\delta \rho \left( \mathbf{%
r}\right) }\right)-\mathbf{\nabla \cdot } \varepsilon \sqrt{\frac{2k_{B}T}{\gamma m}\rho \left( 
\mathbf{r}\right) }\mathbf{\xi }\left( \mathbf{r};t\right)
\end{equation}%
Scaling time by $1/\gamma$ shows that the second-order time derivative is of
higher order in $1/\gamma$ than the other terms so that it can be neglected
in the strong damping approximation leaving 
\begin{equation}
\frac{\partial \rho \left( \mathbf{r}\right) }{\partial t}=\frac{k_{B}T}{%
\gamma m}\mathbf{\nabla \cdot }\left( \rho \left( \mathbf{r}\right) \mathbf{\nabla }%
\frac{\delta \beta F\left[ \rho \right] }{\delta \rho \left( \mathbf{r}%
\right) }\right)-\mathbf{\nabla \cdot }\varepsilon\sqrt{\frac{2k_{B}T}{\gamma m}\rho \left( 
\mathbf{r}\right) }\mathbf{\xi }\left( \mathbf{r};t\right)  \label{DDFT}
\end{equation}%
(Note that this result can be obtained directly from the full hydrodynamic
equations by introducing a scaling of the time and momentum current of $%
t=\gamma t^*$ and $\mathbf{j}=\frac{k_BT}{\gamma l}\bar{\rho}\mathbf{j}^*$, with $\bar{\rho}$ a typical density and $l$ a molecular length scale, and noting that the
dissipative stress tensor is of at least first order in the momentum
current.) Although the coefficient of the noise term is state-dependent, it
is nevertheless the case that the Ito and Stratonovich interpretations of
this stochastic differential equation are the same as shown in Appendix \ref%
{Interpretation}. This model is well-known in the literature as the
Dean-Kawasaki model\cite{Dean,Kawasaki1998} and it has been widely used,
e.g., in the study of the glass transition. Chavanis\cite{Chavanis_Deriv} discusses various approaches to its
interpretation. The derivation sketched here emphasizes the relation between this model and the more general theory of fluctuating hydrodynamics because the latter can be used, in conjunction with the approach described below, to address related problems to which Brownian dynamics is not applicable such as nucleation in simple fluids, multicomponent reacting fluids, etc.. However, for the specific problem of nucleation of over-damped colloids it is  worth noting that the Dean-Kawasaki model can be derived by more rigorous methods as discussed e.g. by Kawasaki\cite{Kawasaki1998} and Chavanis\cite{Chavanis2011}. 

Without the noise term, this is often referred to as Dynamical Density
Functional Theory. Another way of thinking about this is to say that the
usual DDFT results from averaging this result over the noise. The question
then arises as to whether the ``free energy'' entering here is the
thermodynamic free energy or some other quantity (such as a coarse-grained
free energy) that becomes the free energy after averaging over the noise.
From the sketch given above, it would appear consistent to interpret the
free energy functional in Eq.(\ref{DDFT}) as being the free energy
functional for the one-component system of large molecules or colloids since
it arises from a local-equilibrium approximation for the pressure which is
in turn entirely due to interactions between the large-particles.

Note that at low density one has\cite{Evans79} 
\begin{equation}
\beta F\left[ \rho \right] \simeq \int \left( \rho \left( \mathbf{r}\right)
\ln \rho \left( \mathbf{r}\right) -\rho \left( \mathbf{r}\right) \right) d%
\mathbf{r}
\end{equation}%
so that the dynamical equation becomes%
\begin{equation}
\frac{\partial \rho \left( \mathbf{r}\right) }{\partial t}\simeq \frac{k_{B}T%
}{\gamma m}\nabla ^{2}\rho \left( \mathbf{r}\right) -\mathbf{\nabla \cdot }%
\varepsilon \sqrt{\frac{2k_{B}T}{\gamma m}\rho \left( \mathbf{r}\right) }\mathbf{\xi }%
\left( \mathbf{r};t\right)
\end{equation}%
which is simple diffusion with noise. One can therefore identify $D=\frac{%
k_{B}T}{\gamma m}$ as the diffusion constant for the large particles in the
bath so that it, and therefore the coefficient $\gamma$, can be determined
from the Einstein-Stokes equation based solely on the size of the particles
and the properties of the bath.

\subsection{Spherical symmetry}

In principle, this model could be studied numerically. Here, however, the
goal is to make contact with the standard approach to nucleation and one of
the key assumptions usually made is that of spherical symmetry of the
clusters. One cannot, however, simply assume a spherically symmetric density
profile in Eq.(\ref{DDFT}) as the noise term breaks spherical symmetry. One
solution would be to average Eq.(\ref{DDFT}) over a spherical shell but it
proves more convenient and natural to integrate over a spherical volume.

The total (or cumulative) mass inside a spherical shell of radius $r$, $m\left( r\right) $,
is simply the integral of the density%
\begin{equation}
m\left( r\right) =\int_{r^{\prime }<r}\rho \left( \mathbf{r}^{\prime
}\right) d\mathbf{r}^{\prime }.
\end{equation}%
Integrating the dynamical equation over a spherical shell gives an evolution
equation for the cumulative mass%
\begin{eqnarray}
\frac{\partial m\left( r\right) }{\partial t} &=&D\int_{r^{\prime }<r} 
\mathbf{\nabla' \cdot }\left( \rho \left( \mathbf{r}'\right) \mathbf{\nabla' }\frac{%
\delta \beta F\left[ \rho \right] }{\delta \rho \left( \mathbf{r'}\right) }%
\right) d\mathbf{r}^{\prime }-\varepsilon \int_{r^{\prime }<r}\mathbf{\nabla'
\cdot }\sqrt{2D\rho \left( \mathbf{r}^{\prime }\right) }\mathbf{\xi }\left( 
\mathbf{r}^{\prime };t\right) d\mathbf{r}^{\prime } \\
&=&D\int \left( \rho \left( \mathbf{r}\right) \mathbf{\nabla }\frac{\delta
\beta F\left[ \rho \right] }{\delta \rho \left( \mathbf{r}\right) }\right)
\cdot \widehat{\mathbf{r}}d\widehat{\mathbf{r}}-\varepsilon \int_{r^{\prime
}<r}\mathbf{\nabla \cdot }\sqrt{2D\rho \left( \mathbf{r}^{\prime }\right) }%
\mathbf{\xi }\left( \mathbf{r}^{\prime };t\right) d\mathbf{r}^{\prime } 
\notag
\end{eqnarray}%
where Gauss' theorem has been used to convert the first integral on the
right to a surface integral. It is straightforward to show that the
autocorrelation of the noise term is 
\begin{equation}
\delta \left( r-r^{\prime }\right) \delta \left( t-t^{\prime }\right) 2D\varepsilon ^{2}4\pi
r^{2}\left( \frac{1}{4\pi r^{2}}\int \delta \left(
r''-r\right) \rho \left( \mathbf{r''}\right) d\mathbf{r''} \right)
\end{equation}%
where the last term is the density averaged over a spherical shell of radius 
$r$,%
\begin{equation}
<\rho ;R>_{S}\equiv \frac{1}{4\pi R^{2}}\int \delta \left( r-R\right) \rho
\left( \mathbf{r}\right) d\mathbf{r}
\end{equation}%
Hence, the noise term can be replaced by $\varepsilon\sqrt{8\pi Dr^{2}<\rho ;r>_{S}}\xi
(r;t)$ where $<\xi (r;t)\xi (r^{\prime };t^{\prime })>=\delta \left(
r-r^{\prime }\right) \delta \left( t-t^{\prime }\right) $. At this point,
since there is no more coupling to the non-radial noise terms, one can
consistently seek a solution in terms of spherically symmetric density
distributions, $\rho \left( r\right) $, thus giving%
\begin{equation}
\frac{\partial m\left( r\right) }{\partial t}=D4\pi r^{2}\rho \left(
r\right) \frac{\partial }{\partial r}\left. \frac{\delta \beta F\left[ \rho %
\right] }{\delta \rho \left( \mathbf{r}\right) }\right\vert _{\rho \left(
r\right) }-\varepsilon \sqrt{8\pi r^{2}D\rho \left( r\right) }\xi \left(
r;t\right) . \label{genss}
\end{equation}%
It is again easy to verify the equivalence of the Ito and Stratonovich
interpretations of this equation (see Appendix \ref{Interpretation}).

This result can be made somewhat more transparent by noting that for a
spherically symmetric function $\rho \left( \mathbf{r}\right) \equiv \rho \left(
r\right) $ one has%
\begin{equation}
\rho \left( r\right) =\frac{1}{4\pi r^{2}}\int \delta \left( r-r^{\prime
}\right) \rho \left( \mathbf{r}^{\prime }\right) d\mathbf{r}^{\prime }
\end{equation}%
so that for any functional $F$ of $\rho \left( r\right) $ 
\begin{equation}
\frac{\delta F\left[ \rho \right] }{\delta \rho \left( \mathbf{r}\right) }%
=\int_{0}^{\infty }\frac{\delta F\left[ \rho \right] }{\delta \rho \left(
r^{\prime }\right) }\frac{\delta \rho \left( r^{\prime }\right) }{\delta
\rho \left( \mathbf{r}\right) }dr^{\prime }=\frac{\delta F\left[ \rho \right]
}{\delta \rho \left( r\right) }\frac{1}{4\pi r^{2}}
\end{equation}%
giving 
\begin{equation}
\frac{\partial m\left( r\right) }{\partial t}=D4\pi r^{2}\rho \left(
r\right) \frac{\partial }{\partial r}\left( \frac{1}{4\pi r^{2}}\frac{\delta
\beta F\left[ \rho \right] }{\delta \rho \left( r\right) }\right) -\varepsilon 
\sqrt{8\pi r^{2}D\rho \left( r\right) }\xi \left( r;t\right) \label{ss}
\end{equation}%
Next, using the functional chain rule gives%
\begin{eqnarray}
\frac{\partial }{\partial r}\left( \frac{1}{4\pi r^{2}}\frac{\delta \beta F\left[
\rho \right] }{\delta \rho \left( r\right) }\right) &=&\frac{\partial }{%
\partial r}\frac{1}{4\pi r^{2}}\int_{0}^{\infty }\frac{\delta m(r^{\prime })%
}{\delta \rho \left( r\right) }\frac{\delta \beta F\left[ \rho \right] }{\delta
m\left( r^{\prime }\right) }dr^{\prime } \\
&=&\frac{\partial }{\partial r}\frac{1}{4\pi r^{2}}\int_{0}^{\infty }\Theta
\left( r^{\prime }-r\right) 4\pi r^{2}\frac{\delta \beta F\left[ \rho \right] }{%
\delta m\left( r^{\prime }\right) }dr^{\prime }  \notag \\
&=&-\frac{\delta \beta F\left[ \rho \right] }{\delta m\left( r\right) },  \notag
\end{eqnarray}%
where $\Theta(x)$ is the Heaviside step function, so  that the SDE can be written as%
\begin{equation}
\frac{\partial m\left( r\right) }{\partial t}=-D\frac{\partial m\left(
r\right) }{\partial r}\frac{\delta \beta F\left[ \rho \right] }{\delta m\left(
r\right) }-\varepsilon \sqrt{2D\frac{\partial m\left( r\right) }{\partial r}}%
\xi \left( r;t\right).  \label{DynMass}
\end{equation}%
This shows very clearly that the spherically symmetric dynamics is gradient-driven with kinetic coefficients $D\frac{\partial m\left( r\right) }{%
\partial r}$ and obeys a fluctuation-dissipation relation. It also emphasizes
the central role played by the \emph{mass} in this approach to nucleation -
as opposed to density or any order parameter - which is a consequence of the
fact that the theory is founded in hydrodynamics where conservation of mass
is always true.

\subsection{Nucleation pathway}

The stochastic evolution equation, Eq.(\ref{DynMass}), could be used as the
basis of numerical simulations. However, the goal here is to provide a
description of the nucleation pathway and to make connection with more
phenomenological descriptions of the nucleation processes. It is therefore
necessary to characterize the typical or expected pathway. One natural way
to do so is to determine the most likely path (MLP). When the noise is weak,
it can be expected that typical paths will be close to the MLP so that the
latter can be viewed as characterizing the process. If the noise is strong,
the status of the MLP is less clear and it may only be one of many
alternative paths that could be explored by the system.

\subsubsection{The path probability}

The question of the determination of the most likely path for a stochastic
process dates back to Onsager and Machlup\cite{Onsager}. For Gaussian
processes with constant diffusion matrix the result is relatively easy to
derive whereas for the most general cases there are several subtleties and
the most general result seems to have first been derived by Graham\cite%
{Graham}. Here, and in the following sub-subsection the theory of the MLP is
reviewed and then its application to the problem of nucleation is discussed.

Consider the general stochastic dynamics given by%
\begin{equation}
\frac{dx_{i}}{dt}=b_{i}\left( \mathbf{x}\left( t\right) \right) +\epsilon
Q_{ij}\left( \mathbf{x}\left( t\right) \right) \xi _{j}\left( t\right)
\label{StocEq}
\end{equation}%
where the noise terms, $\xi _{j}\left( t\right) $, are Gaussian, white and
uncorrelated. Note that here, and in the following, the Einstein summation
convention is used. Interpreted within the Ito calculus, Graham\cite{Graham} shows that
the probability to observe a given continuous path is given by 
\begin{equation}  \label{Graham1}
P\left( \mathbf{x}\left( t\right) \right) \sim \exp \left(
-\int_{t_{1}}^{t_{N}}\mathcal{L}\left( \mathbf{x}\left( t\right) ,\overset{%
\cdot }{\mathbf{x}}\left( t\right) \right) dt\right)
\end{equation}%
where the Lagrangian is 
\begin{equation}  \label{Graham2}
\mathcal{L}\left( \mathbf{x},\overset{\cdot }{\mathbf{x}}\right) =\frac{1}{2}%
\epsilon ^{-2}\left( \overset{\cdot }{x_{i}}-c_{i}\left( \mathbf{x}\right)
\right) D_{il}^{-1}\left( \mathbf{x}\right) \left( \overset{\cdot }{x_{l}}%
-c_{l}\left( \mathbf{x}\right) \right) +\frac{1}{2}\det \left( Q\left( 
\mathbf{x}\right) \right) \frac{\partial }{\partial x_{i}}\frac{c_{i}\left( 
\mathbf{x}\right) }{\det \left( Q\left( \mathbf{x}\right) \right) }+\epsilon^2 
\frac{R}{12}
\end{equation}%
with%
\begin{eqnarray}  \label{Graham3}
D_{il} &\equiv&Q_{il}Q_{jl} \\
c_{i} &\equiv&b_{i}-\frac{1}{2}\epsilon^{2} \det \left( Q\right) \frac{\partial }{%
\partial x_{j}}\frac{D_{ji}}{\det \left( Q\right) }  \notag,
\end{eqnarray}% 
the Ricci curvature scalar formed using $D_{il}^{-1}\left( \mathbf{x}\left( t\right) \right) $ as a
metric in $x-$space is 
\begin{eqnarray}  \label{Graham4}
R&=&D_{ab}D_{cd}\frac{1}{2}\left(  \frac{\partial^{2}D^{-1}_{ab}}{\partial
x_{c}\partial x_{d}}+\frac{\partial^{2}D^{-1}_{cd}}{\partial x_{a}\partial x_{b}%
}-\frac{\partial^{2}D^{-1}_{bc}}{\partial x_{a}\partial x_{d}}-\frac{\partial
^{2}D^{-1}_{ad}}{\partial x_{b}\partial x_{c}}\right) \notag \\
&& +D_{ab}D_{cd}%
D_{ef}^{-1}\left(  \Gamma_{ab}^{e}\Gamma_{cd}^{f}-\Gamma_{db}^{e}\Gamma
_{ca}^{f}\right) \notag
\end{eqnarray}% 
and the Christoffel symbols of the second kind are
\begin{equation}  \label{Graham5}
\Gamma _{bc}^{a}=\frac{1}{2}D_{ad}\left( \frac{\partial D_{db}^{-1}}{%
\partial x_{c}}+\frac{\partial D_{dc}^{-1}}{\partial x_{b}}-\frac{\partial
D_{bc}^{-1}}{\partial x_{d}}\right).
\end{equation}% 
The probability of observing a path is simply the probability of observing
the required noise needed to generate the path plus additional terms related to the change in variable from the noise to the path. The first term on the right
in the Lagrangian is due to the fact that the probability to observe a
sequence of noises is by definition $P\left( \mathbf{\xi }\left( t\right)
\right) =\exp \left( -\frac{1}{2}\int_{t_{1}}^{t_{N}}\mathbf{\xi }^{2}\left(
t\right) dt\right) $ together with a replacement of $\xi _{j}\left( t\right) 
$ based on rearranging the stochastic differential equation. The second and
third terms are due to the Jacobean in transforming variables from $\mathbf{%
\xi }\left( t\right) $ to $\mathbf{x}\left( t\right) $. Note that the first term is of order $\epsilon ^{-2}$, the second of order $%
\epsilon ^{0}$ and the third of order $\epsilon ^{2}$ so that in the weak
noise limit, only the first term contributes. In this case, another
interpretation of the expression for the path probability is that it is the
Freidlin-Wetzel action functional from the theory of large deviations\cite%
{Freidlin,Varadhan}. The fact that these results can be written in the fully covariant form given above is, as noted by Graham\cite{Graham}, necessary since the path probability must be independent of the choice of variables used to express it. Thus, while the original stochastic model has no explicit geometrical interpretation, the focus of \textit{paths}, which are geometrical objects, induces a natural geometrical perspective in which the (inverse of the) covariance of the noise plays the role of a metric in the space of stochastic variables. 

Given fixed endpoints, the MLP\ is determined by maximizing the probability
with respect to variations in the path resulting in the Euler-Lagrange
equations%
\begin{equation}
\frac{d}{dt}\frac{\partial }{\partial \overset{\cdot }{x_{i}}}\mathcal{L}%
\left( \mathbf{x}\left( t\right) ,\overset{\cdot }{\mathbf{x}}\left(
t\right) \right) -\frac{\partial }{\partial x_{i}}\mathcal{L}\left( \mathbf{x%
}\left( t\right) ,\overset{\cdot }{\mathbf{x}}\left( t\right) \right) =0
\end{equation}%
or, after some manipulations,%
\begin{eqnarray}\label{Dynamic}
\frac{d^{2}}{dt^{2}}x_{r}+\frac{1}{2}\Gamma _{lc}^{r}\frac{dx_{c}}{dt}\frac{%
dx_{l}}{dt}+D_{ri}\left( \frac{\partial D_{cl}^{-1}c_{l}}{\partial x_{i}}-%
\frac{\partial D_{il}^{-1}c_{l}}{\partial x_{c}}\right) \frac{dx_{c}}{dt}%
= \\ D_{ri}\frac{\partial }{\partial x_{i}}\left[ \frac{1}{2}%
c_{m}D_{ml}^{-1}c_{l}+\epsilon ^{2}\frac{1}{2}\det \left( Q\right) \frac{%
\partial }{\partial x_{m}}\frac{c_{m}}{\det \left( Q\right) }+\epsilon ^{4}%
\frac{R}{12}\right]   \notag
\end{eqnarray}%
This can be interpreted as describing a particle with curvilinear
coordinates $\mathbf{x}$ moving in response to a force given by the gradient
of a potential under the metric $\mathbf{D}^{-1}$. (Note that here and below, it is assumed that one selects a solution that actually corresponds to a maximization of the probability, as opposed to a minimum or saddle point extremization.)

Equation(\ref{Dynamic})\ was derived under the condition that the end points
of the path are fixed, so it defines a two-point boundary value problem with
fixed initial point, $\mathbf{x}\left( 0\right) = \mathbf{x}_{0}$, and end point, $\mathbf{x}\left( t_{f}\right) = \mathbf{x}_{f}$
and the time, $t_{f}$, is determined by minimizing the action . In some
cases, the endpoints are not both known \textit{a priori} and one wishes
instead to determine the MLP from any state satisfying some set of
constraints to any other satisfying a different constraints. (For example,
in the nucleation problem, this might be the path connecting a cluster of
total excess mass $M_{1}$ to one with total excess mass $M_{2}$.) If these
constraints involve only the coordinates and are represented in general as $%
K_{a}\left( \mathbf{x}(t_{0})\right) =0$ and $K_{a}\left( \mathbf{x}%
(t_{f})\right) =0$ then the boundary conditions become%
\begin{eqnarray}
\overset{\cdot }{x}_{i}\left( 0\right) &=&c_{i}\left( \mathbf{x}\left(
0\right) \right) +\sum_{a}\lambda _{a}^{\left( 0\right) }D_{ij}^{-1}\left(
x\left( 0\right) \right) \left( \frac{\partial K_{a}\left( x\right) }{%
\partial x_{i}}\right) _{\mathbf{x}\left( 0\right) }  \label{constraints} \\
\overset{\cdot }{x}_{i}\left( t_{f}\right) &=&c_{i}\left( \mathbf{x}\left(
t_{f}\right) \right) +\sum_{a}\lambda _{a}^{\left( f\right)
}D_{ij}^{-1}\left( x\left( t_{f}\right) \right) \left( \frac{\partial
L_{a}\left( x\right) }{\partial x_{i}}\right) _{\mathbf{x}\left(
t_{f}\right) }  \notag
\end{eqnarray}%
where the $\lambda $'s are Lagrange multipliers which must be determined. To
illustrate, suppose the desired path started at point $\mathbf{x}_{0}$ but
that the final point was determined as having a fixed mass, $M\left( \mathbf{%
x}\left( t_{f}\right) \right) =M_{f}$. If there were $n$ parameters, there
would be $n$ constraint equations, Eq.(\ref{constraints}) at the end
point $x_{f}$ together with the condition on the final mass. These $n+1$
constraints would take the place of the $n$ conditions otherwise determined
by specifying the final coordinate as well as providing one additional
constraint for the Lagrange multiplier.

\subsubsection{The most likely path\ when there is a fluctuation-dissipation
relation}

In the limit of weak noise, i.e. for $\epsilon \rightarrow 0$, it can be
shown (see Appendix \ref{proof}) that for for the case of a generalized potential driven dynamics of the form 
\begin{equation}
b_{i}(\mathbf{x})= -L_{ij}(\mathbf{x})\frac{\partial V(\mathbf{x})}{\partial
x_{j}}  \label{a1}
\end{equation}%
that obeys a fluctuation-dissipation relation, 
\begin{equation}
L_{ij}(\mathbf{x})\propto D_{ij}(\mathbf{x})  \label{a2}
\end{equation}%
the MLP between two metastable states, i.e. points satisfying $b_{i}(\mathbf{%
x})=0$, \emph{must} pass through a saddle point and follow the deterministic
path connecting the saddle point to the metastable states, 
\begin{equation}
\frac{dx_{i}}{dt}=\pm L_{ij}(\mathbf{x})\frac{\partial V(\mathbf{x})}{%
\partial x_{j}}  \label{deterministic}
\end{equation}%
It is easily verified by substitution that this is an exact solution to the
general expression, Eq.(\ref{Dynamic}), when Eqs. (\ref{a1}-\ref{a2}) hold
and the initial and final states are metastable points. It can be
interpreted as saying that, starting in the metastable state, one follows
the \emph{time-reversed dynamics} against the free-energy gradient up to
the critical point; beyond the critical point, one follows the forward-time
dynamics down to the stable state. Note that this is only a mathematical
prescription that serves to determine the MLP: there is no suggestion that
the highly dissipative physical system is in any sense time-reversal
invariant. Similarly, the MLP is not the same as the deterministic (i.e. $\epsilon \rightarrow 0$) limit of the stochastic process: it is a separate concept that concerns the most likely path to move both up and down free-energy barriers whereas the deterministic path can only describe movement down a free-energy gradient and is \textit{a priori} incapable of describing movement \textit{up} an energy barrier. While it so happens that for a part of the path between two points that involves moving down a free-energy gradient, the MLP and the deterministic limit coincide, this is not a general identification of the two: thus, while the deterministic path is indeed given by Eq.(\ref{deterministic}) with the minus sign, the more general - and most certainly fluctuation-driven - MLP is given  by Eq.(\ref{deterministic}) with the appropriate sign for the part of the path being determined. 

Recognizing that Eq.(\ref{deterministic}) is equivalent to steepest descent
in  curvilinear coordinates with metric $g_{ij}\equiv L^{-1}_{ij}$  means that
the MLP can be determined using the standard gradient-descent algorithms
\cite{Wales}. This involves first locating the saddle point, $\mathbf{x}_{s}$%
, using, e.g. eigenvector-following techniques\cite{Wales}, and then
perturbing slightly in the direction of the of the unstable eigenvector as
determined from the generalized eigenvalue problem 
\begin{equation}
H_{ij}v_{j}=\lambda L_{ij}^{-1}v_{j}
\end{equation}%
where the Hessian at the saddle point is 
\begin{equation}
H_{ij}=\left. \frac{\partial^2 V(\mathbf{x})}{\partial x_{i}\partial x_{j}}%
\right\vert _{\mathbf{x}_{s}}.
\end{equation}
In principle, the perturbation should be infinitesimal but in practice, some small, finite perturbation must be used with the approximation becoming exact as the size of the perturbation goes to zero. From this initial point, the steepest-descent path is determined using Eq.(\ref{deterministic}) with the positive sign (i.e. the forward-time, deterministic dynamics).

\subsubsection{Application to nucleation}

For the nucleation problem, instead of a collection of variables, $x_{i}$,
as considered above, one has the density $\rho \left( \mathbf{r}\right) $ or
the cumulative mass function, $m\left( r\right) $, which are both fields.
Generally, stochastic differential equations involving fields are understood
to be defined by some sort of spatial discretization scheme so that the
field, say $m\left( r\right) $, is actually a collection of values, $m\left(
r_{i}\right) \equiv m_{i}$, defined on a lattice thus allowing the use of
the results developed for discrete variables.

With this understanding, it is easy to verify that the general model, Eq.(%
\ref{DDFT}), does indeed posses a fluctuation-dissipation relation between
the deterministic and fluctuating parts of the dynamics so that one can
immediately infer that the MLP between the initial and final phases will
pass through a saddle point determined from 
\begin{equation}
\frac{\delta F}{\delta \rho \left( \mathbf{r}\right) }=\mu \label{generalMLP10}
\end{equation}%
where $\mu $ is the chemical potential of the mother phase and will be
determined by moving away from this state in the forward and backward
directions either by following the time-reversed dynamics from the metastable state to the critical state and then the forward-time dynamics thereafter,%
\begin{equation}
\frac{\partial \rho \left( \mathbf{r}\right) }{\partial t}=\pm D\mathbf{%
\nabla \cdot }\rho \left( \mathbf{r}\right) \mathbf{\nabla }\frac{\delta
\beta F\left[ \rho \right] }{\delta \rho \left( \mathbf{r}\right) }, \label{generalMLP20}
\end{equation}%
or, equivalently, by perturbing slightly away from the critical cluster in the direction (or anti-direction) of the most unstable eigenvector and following the forward-time dynamics to the respective minima.
This is the first significant result of this analysis and represents a
generalization of the usual Dynamic Density Functional Theory to barrier
crossing problems. Lacking this result, applications of DDFT have until now
had to rely on the artificial introduction of noise via the initial
conditions of the density in order to induce nucleation (for a recent
example, see Ref. \onlinecite{ArcherEvap}). Such a procedure is in fact
inconsistent in so far as the density appearing in DDFT is already a noise-
and initial-condition-averaged quantity and so, for example, for mother
phases which are fluids it should simply be a constant.

In the case of spherical symmetry, the dynamical model given in Eq.(\ref%
{DynMass}) is obviously of the same form as Eq.(\ref{StocEq}) and obeys the
fluctuation-dissipation constraints of Eqs.(\ref{a1}-\ref{a2}). The MLP in
the limit of weak noise therefore passes through the saddle point determined
from%
\begin{equation}
0=\frac{\delta \beta F\left[ \rho \right] }{\delta m\left( r\right) }=\frac{%
\partial }{\partial r}\left. \frac{\delta \beta F\left[ \rho \right] }{%
\delta \rho \left( \mathbf{r}\right) }\right\vert _{\rho \left( r\right) }
\label{sa}
\end{equation}%
and is given by%
\begin{equation}
\frac{\partial m\left( r\right) }{\partial t}=\pm D\frac{\partial m\left(
r\right) }{\partial r}\frac{\delta \beta F\left[ \rho \right] }{\delta m\left(
r\right) }=\pm D\frac{\partial m\left( r\right) }{\partial r}\frac{\partial 
}{\partial r}\left. \frac{\delta \beta F\left[ \rho \right] }{\delta \rho
\left( \mathbf{r}\right) }\right\vert _{\rho \left( r\right) }  \label{s1}
\end{equation}%
which can be given a geometric interpretation as steepest descent on the
free energy surface in mass-space with a metric of $\left( \frac{\partial
m\left( r\right) }{\partial r}\right) ^{-1}$. Note that this is exactly equivalent to Eqs.(\ref{generalMLP10}-\ref{generalMLP20}) with the assumption of spherical symmetry, thus showing the equivalence of the approaches. It also  allows one to define the
distance between two mass distributions, $m_{0}\left( r\right) $ and $%
m_{1}\left( r\right) $, as%
\begin{equation}
s\left[ m_{0},m_{1}\right] =\min_{T}\min_{\mathrm{paths}}\int_{0}^{T}\sqrt{%
\int_{0}^{\infty }\frac{\partial m\left( r,t\right) }{\partial t}\left( 
\frac{\partial m\left( r,t\right) }{\partial r}\right) ^{-1}\frac{\partial
m\left( r,t\right) }{\partial t}dr}dt  \label{s0}
\end{equation}%
where the right hand side is minimized over all "acceptable" paths
connecting the two mass distributions with $m(r,0)=m_{0}(r)$ and $%
m(r,T)=m_{1}(r)$. An "acceptable" path is one obeying the basic constraints
on the cumulative mass distribution that it be monotonically increasing as a
function of $r$ and that $m(0)=0$. These properties mean that the position $%
r $ may be replaced by $m(r)$ as the variable of integration yielding 
\begin{eqnarray}  \label{genmetric}
s\left[ m_{0},m_{1}\right] &=&\min_{T}\min_{\mathrm{paths}}\int_{0}^{T}\sqrt{%
\int_{0}^{\infty }\left( \frac{\partial m\left( r,t\right) }{\partial t}/%
\frac{\partial m\left( r,t\right) }{\partial r}\right) ^{2}\frac{\partial
m\left( r,t\right) }{\partial r}dr}dt \\
&=&\min_{T}\min_{\mathrm{paths}}\int_{0}^{T}\sqrt{\int_{0}^{\infty }\left( \frac{%
\partial r\left( m,t\right) }{\partial t}\right) ^{2}dm}dt  \notag \\
&=&\sqrt{\int_{0}^{\infty }\left( m_{0}^{-1}\left( x\right)
-m_{1}^{-1}\left( x\right) \right) ^{2}dx}  \notag
\end{eqnarray}%
where the last line follows from recognizing that the second line now
defines a Euclidean metric. Since $m_{0}^{-1}\left( x\right) $ is the
location at which distribution $m_{0}\left( r\right) $ has value $x$ and $%
m_{1}^{-1}\left( x\right) $ is the analogous quantity for the second
distribution, this expresses the distance between the distributions in terms
of the sum over Euclidean distances between equal-mass points. This
structure is entirely due to the presence of the factor $\frac{\partial
m\left( r,t\right) }{\partial r}\sim \rho \left( r\right) $ in the dynamical
equations and, via the fluctuation-dissipation relation, in the noise.
Without it, Eq.(\ref{s0}) would already be Euclidean and the distance
function would simply depend on the integral over $\left( m_{0}\left(
r\right) -m_{1}\left( r\right) \right) ^{2}$.

The fact that the exact distance function can be easily calculated means
that powerful approximate methods for determining the steepest descent path,
such as the String method\cite{String} and the Nudged Elastic Band\cite%
{NEB,NEB0}, can be used to determine the steepest descent paths defined by
Eq.(\ref{s1}) which is to say, the MLP. Another implication concerns the
stochastic evolution equations themselves. The fact that the induced
distance function is Euclidean in the variable $r(m)$, immediately implies
that if the SDE\ is written in terms of this variable, the noise will be
white (i.e. the coefficient of the noise variable $\xi(r;t)$ will be constant rather than state-dependent). That this is the case is easily verified by first observing that%
\begin{equation}
\frac{\partial r\left( m\right) }{\partial t}=\frac{\partial r}{\partial m}%
\frac{\partial m}{\partial t}=\left( \frac{\partial m}{\partial r}\right)
^{-1}\frac{\partial m}{\partial t}
\end{equation}%
Using the evolution equation, Eq.(\ref{DynMass}), gives%
\begin{equation}
\frac{\partial r\left( m\right) }{\partial t}=\left( \frac{\partial m}{%
\partial r}\right) ^{-1}\frac{\partial m\left( r\right) }{\partial t}=-D%
\frac{\delta \beta F\left[ \rho \right] }{\delta m\left( r\right) }-\epsilon \sqrt{%
2D\left( \frac{\partial m\left( r\right) }{\partial r}\right) ^{-1}}\xi
\left( r;t\right)
\end{equation}%
where the term $\frac{\delta F\left[ \rho \right] }{\delta m\left( r\right) }
$ must be expressed in terms of $r(m)$. At the moment, it does not seem as
if the goal of getting white noise is going to be achieved. However, note
that one must express the noise in terms of $m$ as the dependent variable
rather than $r$. Since%
\begin{eqnarray}
\left\langle \xi \left( r;t\right) \xi \left( r^{\prime };t^{\prime }\right)
\right\rangle &=&\delta \left( r-r^{\prime }\right) \delta \left(
t-t^{\prime }\right) \\
&= &\delta \left( r\left( m\right) -r\left( m^{\prime }\right)
\right) \delta \left( t-t^{\prime }\right)  \notag \\
&=&\left( \frac{\partial r\left( m\right) }{\partial m}\right) ^{-1}\delta
\left( m-m^{\prime }\right) \delta \left( t-t^{\prime }\right)  \notag
\end{eqnarray}%
it is clear that one can make the substitution $\xi \left( r;t\right)
\longmapsto \sqrt{\left( \frac{\partial r\left( m\right) }{\partial m}%
\right) ^{-1}}\xi \left( m;t\right) $ giving%
\begin{equation}
\frac{\partial r\left( m\right) }{\partial t}=-D\left. \frac{\delta \beta F\left[
\rho \right] }{\delta m\left( r\right) }\right\vert _{r\left( m\right)
}-\epsilon \sqrt{2D}\xi \left( m;t\right)
\end{equation}%
as expected. It is not clear whether this formulation is of any practical
significance.

\section{Order parameters}

So far, the central quantity governing the description of nucleation has
been the cumulative mass, $m(r;t)$. However, the classical description of
nucleation is typically formulated in terms of a small number, often just
one, order parameter. For example, in Classical Nucleation Theory clusters
of the new phase are assumed to be spherical, with a radius $R$, to have the
bulk new phase in the interior of the cluster and to have an interface of
vanishing width. The only quantity that can vary is therefore the radius and
this is the order parameter in terms of which the dynamics are formulated. 

The point of view adopted here in order to make a connection with the
concept of order parameters is that the parameters must determine an
approximation to the cumulative mass distribution or, equivalently, to the
density. For example, in the CNT picture described above, a cluster has
interior density $\rho _{0}$ which is just the bulk density of the new
phase; exterior density $\rho _{\infty }$ which is the density of the
metastable phase, a radius $R$ and an interface of negligible width. Hence,
it implicitly specifies a density model%
\begin{equation}
\rho \left( r;t\right) \longmapsto \rho \left( r;R(t)\right) \equiv \rho _{0}\Theta \left( R(t)-r\right) +\rho _{\infty
}\Theta \left( r-R(t)\right)
\end{equation}%
and a cumulative mass of%
\begin{equation}
m\left( r;t \right) \longmapsto  m\left( r; R(t)\right) \equiv \frac{4\pi }{3}r^{3}\rho _{0}\Theta \left( R(t)-r\right) +%
\frac{4\pi }{3}\left( R^{3}(t)\rho _{0}+\left( r^{3}-R^{3}(t)\right) \rho _{\infty
}\right) \Theta \left( r-R(t)\right)
\end{equation}%
In general, the approximating function can depend on many parameters so that the general case is $\rho \left( r;t\right) \longmapsto \rho \left( r;\mathbf{x}(t)\right)$ for a set of time-dependent parameters $\mathbf{x}(t)$. The goal in this Section is to determine how the time-dependence of the restricted description in terms of the order parameters can be developed from the general theory. 
The reason for investigating this question is that it has been shown that one can derive the CNT description of the critical cluster as
the lowest-order result in a systematic expansion of the DFT free energy based on these types of
parameterizations of the density\cite{LutskoBubble1,Lutsko2011a} so that this gives a path for making contact between the general theory and CNT.
Finally, it is also relevant to note that any practical scheme for the
integration of Eq.(\ref{DynMass}) will involve some sort of
representation of the field in terms of a finite number of parameters, so
the results obtained here will be of equal use in numerical calculations as
discussed below.

In the following, two different approaches to the use of order parameters
will be described. The first is solely concerned with the determination of
the most likely path for nucleation and involves relatively few assumptions.
The second is aimed at determining approximate stochastic equations for the
order parameters which, in turn, can be used to determine the MLP. The
approaches have different strengths and weaknesses: the former involves
fewer assumptions but is more restricted in its use while the latter gives
broader contact with earlier work but is somewhat heuristic. Both methods
become exact in the limit of a complete set of order parameters and of
course agree with one another in that limit, as demonstrated below.

\subsection{Direct determination of pathway using order parameters}

In the weak noise limit, the probability for a path based on the
spherically-symmetric dynamics, Eq.(\ref{ss}), is 
\begin{equation}
P\sim \exp \left( -\frac{1}{8\pi D \varepsilon}\int_{0}^{T}\mathcal{L}dt\right)
\end{equation}%
with Lagrangian%
\begin{equation}
\mathcal{L}=\frac{1}{2}\int_{0}^{\infty }\frac{1}{ r^{2}\rho \left(
r\right) }\left( \frac{\partial m\left( r\right) }{\partial t}-Dr^{2}\rho
\left( r\right) \frac{\partial }{\partial r}\frac{1}{r^2}\frac{\delta \beta F\left[ \rho \right]
}{\delta \rho \left( r\right) }\right) ^{2}dr  \label{Lagrangian}
\end{equation}%
It is shown in Appendix \ref{AppExact} that if the density is parametrized
as $\rho (r;t)=\rho (r;\mathbf{x}(t))$, then the weak-noise Lagrangian becomes%
\begin{equation}
\mathcal{L}=\frac{1}{2}g_{ab}\left( \mathbf{x}\right) \frac{dx_{a}}{dt}\frac{%
dx_{b}}{dt}+D\frac{dx_{a}}{dt}\frac{\partial \beta \Omega \left[
\rho \right] }{\partial x_{a}}+V\left( \mathbf{x}\right)
\end{equation}%
where $\Omega[\rho]=F[\rho]-\mu \rho$ and %
\begin{equation}
V\left( \mathbf{x}\right) =\frac{1}{2}D^{2}\int_{0}^{\infty }r^{2}\rho
\left( r\right) \left( \frac{\partial }{\partial r}\frac{1}{r^2}\frac{\delta \beta F\left[ \rho %
\right] }{\delta \rho \left( r\right) }\right) ^{2}dr
\end{equation}%
and where the metric $g_{ab}\left( \mathbf{x}\right) $ is 
\begin{equation}
g_{ab}=\int_{0}^{\infty} \frac{1}{4\pi r^2 \rho(r)} \frac{\partial m(r)}{%
\partial x_a}\frac{\partial m(r)}{\partial x_b}dr \label{metric1}
\end{equation}

Maximizing the path probability gives the Euler-Lagrange equations%
\begin{equation}
g_{il}\frac{d^{2}x_{l}}{dt^{2}}+\frac{1}{2}\frac{dx_{l}}{dt}\frac{dx_{j}}{dt}%
\left( \frac{\partial g_{il}}{\partial x_{j}}+\frac{\partial g_{ij}}{%
\partial x_{l}}-\frac{\partial g_{lj}}{\partial x_{i}}\right) =
\frac{\partial }{\partial x_{i}}V(\mathbf{x}) \label{general0}
\end{equation}%
or%
\begin{equation}
\frac{d^{2}x_{i}}{dt^{2}}+\Gamma _{lj}^{i}\frac{dx_{l}}{dt}\frac{dx_{j}}{dt}%
=g_{ir}^{-1}\frac{\partial }{\partial x_{r}}V(\mathbf{x}) \label{general}
\end{equation}%
where the Christoffel symbol of the second kind is 
\begin{equation}
\Gamma _{kl}^{i}=\frac{1}{2}g_{im}^{-1}\left( \frac{\partial g_{mk}}{%
\partial x_{l}}+\frac{\partial g_{ml}}{\partial x_{k}}-\frac{\partial g_{kl}%
}{\partial x_{m}}\right) .
\end{equation}%
Equation (\ref{general}) is recognized as the equation of motion of a
particle expressed in curvilinear coordinates with a force derived from the
potential $V(\mathbf{x})$. If the end points are determined by the constraints, 
\begin{eqnarray}
J_{l}\left( \mathbf{x}\left( 0\right) \right)  &=&0,\;l=1,...,n \\
K_{l}\left( \mathbf{x}\left( T\right) \right)  &=&0,\;l=1,...,n  \notag
\end{eqnarray}%
then the boundary conditions are 
\begin{eqnarray}
\left. \frac{dx_{i}}{dt}\right\vert _{t=0}+\frac{D}{4\pi }\left.
g^{il}\left( \mathbf{x}\right) \frac{\partial \beta F\left[ \rho \right] }{%
\partial x_{l}}\right\vert _{t=0} &=&\left. \mu _{r}g^{il}\left( \mathbf{x}%
\right) \frac{\partial J_{r}}{\partial x_{l}}\right\vert _{t=0} \\
\left. \frac{dx_{i}}{dt}\right\vert _{t=T}+\frac{D}{4\pi }\left.
g^{il}\left( \mathbf{x}\right) \frac{\partial \beta F\left[ \rho \right] }{%
\partial x_{l}}\right\vert _{t=T} &=&\left. \lambda _{r}g^{il}\left( \mathbf{%
x}\right) \frac{\partial K_{r}}{\partial x_{l}}\right\vert _{t=T}  \notag
\end{eqnarray}%
where $\mu _{r}$ and $\lambda _{r}$ are Lagrange multipliers introduced in
the course of minimizing the action under the constraints. Note that if,
e.g., the initial point is fixed, then $J_{r}=x_{r}\left( T\right) -x_{0r}$
and the constraint just serves to evaluate the (uninteresting) Lagrange
multiplier: in this case the constraint equations play no role. On the
other hand, if a nontrivial constraint is applied then instead of fixing the
value of $x_{r}\left( T\right) $ a condition is imposed on the velocities at
time $T$. For example, for nucleation one might know the initial state, $%
\mathbf{x}\left( 0\right) $, and might want to know the most likely final
state with a given excess mass, 
\begin{equation}
M=4\pi \int_{0}^{\infty }\left( \rho \left( r;\mathbf{x}\left( T\right)
\right) -\rho _{\infty }\right) r^{2}dr.
\end{equation}%
Then, the boundary conditions at the end point would be 
\begin{equation}
\left. \frac{dx_{i}}{dt}\right\vert _{t=T}+\frac{D}{4\pi }\left.
g_{il}^{-1}\left( \mathbf{x}\right) \frac{\partial \beta F\left[ \rho \right] }{%
\partial x_{l}}\right\vert _{t=T}=\lambda g_{il}^{-1}\left( \mathbf{x}\left(
T\right) \right) 4\pi \int_{0}^{\infty }\frac{\partial \rho \left( r;\mathbf{%
x}\left( T\right) \right) }{\partial x_{l}\left( T\right) }r^{2}dr
\end{equation}

\subsection{An order-parameter dynamics} \label{opd}

The spatial and temporal components of the density can be separated by
writing it in some parametrized form as $\rho \left( r;t\right) =\rho \left(
r;\mathbf{x}\left( t\right) \right) $ where $\mathbf{x}\left( t\right) $ is
a collection of scalar parameters. One common method of doing so would be to
expand the density in a complete set of basis functions. Alternatively, the
parameters could represent a discretization of the density, $x_{n}=\rho
\left( n\Delta \right) $ for some small distance $\Delta $. As these examples
indicate, there will be in general an arbitrary number of parameters. Their
time evolution can be developed by noting that the $m(r)$ will also be a
function of the parameters so that from Eq.(\ref{genss}) one can write%
\begin{equation}
\frac{\partial m\left( r;\mathbf{x}\left( t\right) \right) }{\partial x_{i}}%
\frac{dx_{i}}{dt}=D4\pi r^{2}\rho \left( r;\mathbf{x}\left( t\right) \right) 
\frac{\partial }{\partial r}\left. \frac{\delta \beta F\left[ \rho \right] }{%
\delta \rho \left( \mathbf{r}\right) }\right\vert _{\rho \left( r;\mathbf{x}%
\left( t\right) \right) }-\epsilon \sqrt{8\pi r^{2}D\rho \left( r;\mathbf{x}%
\left( t\right) \right) }\xi \left( r;t\right) .  \label{begin}
\end{equation}%
Suppose that this is multiplied by some function $W_{j}\left( r;\mathbf{x}%
\left( t\right) \right) $ and integrated giving 
\begin{eqnarray}
g_{ij}\left( \mathbf{x}\right) \frac{dx_{i}}{dt} &=&D\int_{0}^{\infty
}W_{j}\left( r;\mathbf{x}\left( t\right) \right) 4\pi r^{2}\rho \left( r;%
\mathbf{x}\left( t\right) \right) \frac{\partial }{\partial r}\left( \frac{%
\delta \beta F\left[ \rho \right] }{\delta \rho \left( \mathbf{r}\right) }%
\right) _{\rho \left( r;\mathbf{x}\left( t\right) \right) }dr  \label{E1} \\
&-&\int W_{j}\left( r;\mathbf{x}\left( t\right) \right) \epsilon \sqrt{8\pi
r^{2}D\rho \left( r;\mathbf{x}\left( t\right) \right) }\xi \left( r;t\right)
dr.  \notag
\end{eqnarray}%
where 
\begin{equation}
g_{ij}\left( \mathbf{x}\right) =\int_{0}^{\infty }W_{i}\left( r;\mathbf{x}%
\left( t\right) \right) \frac{\partial m\left( r;\mathbf{x}\left( t\right)
\right) }{\partial x_{j}}dr
\end{equation}%
The autocorrelation of the noise term is easily calculated giving 
\begin{equation}
\mathcal{\epsilon }^{2}2D\int_{0}^{\infty }4\pi r^{2}W_{i}\left( r;\mathbf{x}%
\left( t\right) \right) W_{j}\left( r;\mathbf{x}\left( t\right) \right) \rho
\left( r;\mathbf{x}\left( t\right) \right) dr
\end{equation}%
While these equations are exact regardless of the choice of $W_{i}\left( r;%
\mathbf{x}\left( t\right) \right) $, it will soon be apparent the the result
is particularly simple if the diffusion matrix and the matrix $g_{ij}\left( 
\mathbf{x}\right) $ are required to be proportional which is only possible
if, modulo a multiplicative constant, 
\begin{equation}
W_{i}\left( r;\mathbf{x}\left( t\right) \right) =\frac{1}{4\pi r^{2}\rho
\left( r;\mathbf{x}\left( t\right) \right) }\frac{\partial m\left( r;\mathbf{%
x}\left( t\right) \right) }{\partial x_{i}}
\end{equation}%
so that $\mathcal{D}_{ij}\left( \mathbf{x}\right) =2Dg_{ij}\left( \mathbf{x}%
\right) $ with
\begin{equation}
g_{ij}\left( \mathbf{x}\right) =\int_{0}^{\infty }\frac{1}{4\pi r^{2}\rho
\left( r;\mathbf{x}\left( t\right) \right) }\frac{\partial m\left( r;\mathbf{%
x}\left( t\right) \right) }{\partial x_{i}}\frac{\partial m\left( r;\mathbf{x%
}\left( t\right) \right) }{\partial x_{j}}dr,  \label{metric}
\end{equation}%
i.e., the same metric as found above (see Eq.(\ref{metric1})). Another result of this choice is that the thermodynamic driving force becomes%
\begin{eqnarray}
&&\int_{0}^{\infty }W_{j}\left( r;\mathbf{x}\left( t\right) \right) 4\pi
r^{2}\rho \left( r;\mathbf{x}\left( t\right) \right) \frac{\partial }{%
\partial r}\left. \frac{\delta \beta F\left[ \rho \right] }{\delta \rho
\left( \mathbf{r}\right) }\right\vert _{\rho \left( r;\mathbf{x}\left( t\right)
\right) }dr  \label{parts} \\
&=&\int_{0}^{\infty }\frac{\partial m\left( r;\mathbf{x}\left( t\right)
\right) }{\partial x_{i}}\frac{\partial }{\partial r}\left. \frac{\delta
\beta F\left[ \rho \right] }{\delta \rho \left( \mathbf{r}\right) }\right\vert
_{\rho \left( r;\mathbf{x}\left( t\right) \right) }dr  \notag \\
&=&\left[ \frac{\partial m\left( r;\mathbf{x}\left( t\right) \right) }{%
\partial x_{i}}\left. \frac{\delta \beta F\left[ \rho \right] }{\delta \rho
\left( \mathbf{r}\right) }\right\vert _{\rho \left( r;\mathbf{x}\left( t\right)
\right) }\right] _{0}^{\infty }-\int_{0}^{\infty }\frac{\partial ^{2}m\left(
r;\mathbf{x}\left( t\right) \right) }{\partial x_{i}\partial r}\left. \frac{%
\delta \beta F\left[ \rho \right] }{\delta \rho \left( \mathbf{r}\right) }%
\right\vert _{\rho \left( r;\mathbf{x}\left( t\right) \right) }dr  \notag \\
&=&\left[ \frac{\partial m\left( r;\mathbf{x}\left( t\right) \right) }{%
\partial x_{i}}\left. \frac{\delta \beta F\left[ \rho \right] }{\delta \rho
\left( \mathbf{r}\right) }\right\vert _{\rho \left( r;\mathbf{x}\left( t\right)
\right) }\right] _{0}^{\infty }-\int \frac{\partial \rho \left( r;\mathbf{x}%
\left( t\right) \right) }{\partial x_{i}}\left. \frac{\delta \beta F\left[
\rho \right] }{\delta \rho \left( \mathbf{r}\right) }\right\vert _{\rho \left( r;%
\mathbf{x}\left( t\right) \right) }d\mathbf{r}  \notag
\end{eqnarray}%
For the problem of nucleation, the first term gives no contribution at $r=0$
while for very large $r$, one expects the system to have the bulk properties
giving 
\begin{equation}
\lim_{r\rightarrow \infty }\frac{\partial m\left( r;\mathbf{x}\left(
t\right) \right) }{\partial x_{i}}\left. \frac{\delta \beta F\left[ \rho %
\right] }{\delta \rho \left( \mathbf{r}\right) }\right\vert _{\rho \left( r;%
\mathbf{x}\left( t\right) \right) }=\frac{\partial N\left( \mathbf{x}\left(
t\right) \right) }{\partial x_{i}}\mu
\end{equation}%
where $N$ is the total number of particles and $\mu $ is the chemical
potential in the bulk. Using the functional chain rule, the second term is
recognized as 
\begin{equation}
\int \frac{\partial \rho \left( r;\mathbf{x}\left( t\right) \right) }{%
\partial x_{i}}\left. \frac{\delta \beta F\left[ \rho \right] }{\delta \rho
\left( \mathbf{r}\right) }\right\vert _{\rho \left( r;\mathbf{x}\left( t\right)
\right) }d\mathbf{r=}\frac{\partial \beta F\left( \mathbf{x}\right) }{%
\partial x_{i}}
\end{equation}%
where $F\left( \mathbf{x}\right) =F\left[ f\right] $. (Note that this is
only exact if the parametrization is complete.) Combining these results
gives the stochastic model%
\begin{equation}
g_{ij}\left( \mathbf{x}\right) \frac{dx_{j}}{dt}=-D\frac{\partial \beta
\Omega }{\partial x_{i}}-\epsilon \int \sqrt{\frac{2D}{4\pi r^{2}\rho \left(
r;\mathbf{x}\left( t\right) \right) }}\frac{\partial m\left( r;\mathbf{x}%
\left( t\right) \right) }{\partial x_{i}}\xi \left( r;t\right) dr.
\end{equation}%
where again $\Omega =F-\mu N$. Ideally, one would like to replace the noise by a
simpler form giving the same autocorrelation function however there is a
complication. The use of the usual chain rule for derivatives means that the
resulting stochastic differential equation must be understood in the
Stratonovich interpretation\cite{Gardiner} and, unfortunately, in the
present case the resulting equation is not Ito-Stratonovich equivalent. It
is therefore not the case that one can substitute one noise term for another
with the same autocorrelation matrix as this only holds in the Ito form. As
described in Appendix \ref{Noise}, the spurious drift arising from Ito-Stratonovich inequivalence  gives rise to an additional
contribution to the deterministic part of the equation with the final result%
\begin{equation}
\frac{dx_{j}}{dt}=-Dg_{ij}^{-1}\left( \mathbf{x}\right) \frac{\partial \beta
\Omega }{\partial x_{i}}+2D\epsilon ^{2}A_{i}\left( \mathbf{x}\right)
-\epsilon \sqrt{2D}q_{ji}^{-1}\left( \mathbf{x}\right) \xi _{j}\left(
t\right)  \label{final}
\end{equation}%
with $g_{ij}=q_{il}q_{jl}$ (note that $g$ is positive semi-definite so this
decomposition is always possible) and what will be termed the ``anomalous force''  (since it does not arise from the thermodynamic driving force) is
\begin{eqnarray}
A_{i}\left( \mathbf{x}\right) &=&q_{ik}^{-1}\left( \mathbf{x}\right) \frac{%
\partial q_{jk}^{-1}\left( \mathbf{x}\right) }{\partial x_{j}}-\frac{1}{2}%
g_{il}^{-1}\left( \mathbf{x}\right) \frac{\partial g_{jm}^{-1}\left( \mathbf{%
x}\right) }{\partial x_{l}}g_{mj}\left( \mathbf{x}\right) \\
&&+\frac{1}{2}\left( g_{il}^{-1}\left( \mathbf{x}\right) g_{jm}^{-1}\left( 
\mathbf{x}\right) -g_{ij}^{-1}\left( \mathbf{x}\right) g_{lm}^{-1}\left( 
\mathbf{x}\right) \right) \int_{0}^{\infty }\frac{1}{4\pi r^{2}\rho
^{2}\left( r;\mathbf{x}\right) }\frac{\partial \rho \left( r;\mathbf{%
x}\right) }{\partial x_{l}}\frac{\partial m\left( r;\mathbf{x}%
\right) }{\partial x_{j}}\frac{\partial m\left( r;\mathbf{x}\right) }{%
\partial x_{m}}dr  \notag
\end{eqnarray}%
The presence of the anomalous force means that there is no exact
fluctuation-dissipation relation. However, since it has its origin in the
noise, and is therefore of order $\epsilon ^{2}$, it will not affect the
weak-noise limit and the MLP\ will be determined as usual from the forward-
and backward-time dynamics starting at the critical point. Again, it is
important to remember that in the strong-noise regime Eq. (\ref{final}) must
be understood in the Stratonovich interpretation :\ the equivalent Ito
equation will have a modified form of the anomalous force (see Appendix \ref%
{Noise} for details).

The advantages of the choice for $F_{i}\left( r;\mathbf{x}\left( t\right)
\right) $ are now evident. First, the particular choice used here results in
the free-energy term having the form of a simple gradient of the free energy
due to the ability to integrate by parts in Eq.(\ref{parts}). Second, the
fact that the fluctuation-dissipation relation is preserved means that in
the weak-noise limit one recovers the classical behavior with the MLP
passing through the critical cluster.

Equation (\ref{final}) is exact when the parameters are determined from a
complete representation of the density. For a \emph{finite} collection of
parameters, it seems reasonable to continue to use Eq.(\ref{final}) as an
approximation and numerical evidence in support of this will be given below.
Appendix \ref{AltDerivation} gives an alternative derivation of this model
showing that it is reasonable even for an incomplete (i.e. approximate)
representation of the density provided that the difference between $\rho
\left( r;\mathbf{x}\left( t\right) \right) $ and the actual density $\rho
\left( r;t\right) $ is in some sense small.

\subsection{Most likely path}

A heuristic stochastic dynamics for a set of order parameters is given in
Eq.(\ref{final}) above. In the weak noise limit, in which the anomalous force
 can be neglected, this dynamics is gradient-driven and satisfies a fluctuation-dissipation relation
(FDR)  so that it immediately follows that the MLP connecting the two
metastable phases passes through the saddle point defined by%
\begin{equation}  \label{g0}
0=\frac{\partial \beta \Omega }{\partial x_{i}}
\end{equation}%
and is determined by steepest-descent according to%
\begin{equation}
\frac{dx_{i}}{dt}= - Dg_{ij}^{-1}\left( \mathbf{x}\right) \frac{\partial
\beta \Omega }{\partial x_{j}}  \label{g1}
\end{equation}%
Following the same arguments given above, the general MLP connecting any two
states is determined from the Euler-Lagrange equations 
\begin{equation}
g_{ij}\frac{d^{2}x_{j}}{dt^{2}}+\frac{1}{2}\left( \frac{\partial g_{is}}{%
\partial x_{r}}+\frac{\partial g_{ir}}{\partial x_{s}}-\frac{\partial g_{rs}%
}{\partial x_{i}}\right) \frac{dx_{r}}{dt}\frac{dx_{s}}{dt}=\frac{\partial }{%
\partial x_{i}}\left( \frac{D^2}{2}\frac{\partial \beta \Omega }{\partial x_{j}%
}g_{jl}^{-1}\frac{\partial \beta \Omega }{\partial x_{l}}\right)  \label{g2}
\end{equation}%
It is easy to show by direct substitution that Eq.(\ref{g1}) is a solution
to Eq. (\ref{g2}).

Two expressions for the evolution of order parameters have been derived:\
Eq.(\ref{g1}), which follows from the general equations for the MLP, Eq.(\ref%
{g2}), for the special case that the endpoints are metastable states, and
Eq.(\ref{general0}), which follows from the minimization of the path
probability after assuming a particular test function. In essence, the
former results from introducing the approximate density after determining
the equations for the MLP while the latter results from introducing the
approximate density before determining the MLP. Each formulation has its
advantages:\ the gradient formulation shares certain properties with the
exact MLP such as the role of the critical cluster whereas it is not even
clear a priori that Eq.(\ref{general0}) will give a path passing through the
critical point. On the other hand, Eq.(\ref{general0}) is exact, given the
assumed form for the density, whereas the gradient equations are derived via
a series of manipulations which are not unique: one could derive any number
of equations of motion for the parameters from Eq.(\ref{begin}) by similar
manipulations to those used above. It is therefore of some interest to
examine the connection between these two formulations.

Comparing Eq.(\ref{general}) and Eq.(\ref{g2}) it is clear that the only
difference is in the source terms and that equivalence would demand that%
\begin{equation}
\int_0^{\infty} r^{2}\rho \left( r\right) \left( \frac{\partial }{\partial r}\frac{1}{r^2}\frac{%
\delta \beta F\left[ \rho \right] }{\delta \rho \left( r\right) }\right)
^{2}dr=\frac{\partial \beta \Omega }{\partial x_{j}}g_{jl}^{-1}\frac{%
\partial \beta \Omega }{\partial x_{l}}  \label{g3}
\end{equation}%
Now, from the derivations above, it is already known that%
\begin{equation}
\frac{\partial \beta \Omega }{\partial x_{j}}=-\int_0^{\infty} \frac{\partial m\left(
r\right) }{\partial x_{j}}\frac{\partial }{\partial r}\frac{1}{r^2}\frac{\delta \beta F%
\left[ \rho \right] }{\delta \rho \left( r\right) }dr
\end{equation}%
so%
\begin{equation}
\frac{\partial \beta \Omega }{\partial x_{j}}g_{jl}^{-1}\frac{\partial \beta
\Omega }{\partial x_{l}}=\int_0^{\infty} dr^{\prime }\int_0^{\infty} drr^{2}\rho \left( r\right)
\left( \frac{\partial }{\partial r}\frac{1}{r^2}\frac{\delta \beta F\left[ \rho \right] }{%
\delta \rho \left( r\right) }\right) \left( \frac{\partial }{\partial
r^{\prime }}\frac{1}{r'^2}\frac{\delta \beta F\left[ \rho \right] }{\delta \rho \left( r^{\prime
}\right) }\right) \left( r^{-2}\rho ^{-1}\left( r\right) \frac{\partial
m\left( r\right) }{\partial x_{j}}g_{jl}^{-1}\frac{\partial m\left(
r^{\prime }\right) }{\partial x_{l}}\right)  \notag
\end{equation}%
which is the same as Eq.(\ref{g3}) provided that 
\begin{equation}
\frac{\partial m\left( r\right) }{\partial x_{i}}g_{ij}^{-1}\frac{\partial
m\left( r^{\prime }\right) }{\partial x_{j}}=r^{2}\rho \left( r\right)
\delta \left( r-r^{\prime }\right) .  \label{complete}
\end{equation}%
As demonstrated in the Appendix \ref{AppComplete}, this is a completeness
relation and is sufficient to give equality between the two descriptions of
the MLP. It is satisfied when the parameters result from an expansion of the
density in a complete set of basis functions. The use of the order-parameter
dynamics for a \emph{finite} collection of order parameters is therefore
heuristic. However, as shown below, only a small number of parameters is
necessary for the heuristic dynamics to show convergence  to the exact result.

\section{Applications}

The classic problem of the nucleation of a liquid droplet in a
supersaturated vapor will be used to illustrate the theory described above.
Calculations were performed using the squared-gradient free energy model, 
\begin{equation}
F[\rho ]=\int \left( f(\rho (\mathbf{r}))+\frac{1}{2}K(\nabla \rho (\mathbf{r%
}))^{2}\right) d\mathbf{r}
\end{equation}%
where $f(\rho )$ is the bulk free energy per unit volume. The calculations
reported here were performed for the ten Wolde-Frenkel model potential for
globular proteins\cite{tWF-Proteins} which consists of a hard core of radius 
$\sigma $ and an effective pair potential outside the core,%
\begin{equation}
v\left( r>\sigma \right) =\frac{4\,\epsilon }{\alpha ^{2}}\left( \,\left( 
\frac{1}{(\frac{r}{\sigma })^{2}-1}\right) ^{6}-\,\alpha \,\left( \frac{1}{(%
\frac{r}{\sigma })^{2}-1}\right) ^{3}\right)
\end{equation}%
where $\epsilon $ is the energy scale of the potential and the parameter $%
\alpha $ controls its range. For large values of $\alpha $, such as $\alpha
=50$ used here, this gives a phase diagram typical of proteins including a
metastable high-concentration ("liquid") phase\cite%
{GuntonBook,lutsko:244907}. The results presented here pertain to the
transition from the low-concentration ("gas") phase to a high-concentration
("liquid")\ phase which is known to sometimes play a role as the first step
in the processes of protein crystallization\cite%
{VekilovCGDReview2004,GuntonBook}. The bulk Helmholtz free-energy function, $%
f\left( \rho \right) $, was approximated using first-order perturbation
theory\cite{BarkerHend,HansenMcdonald} as described in detail in Refs. %
\onlinecite{lutsko:244907,protein}. The coefficient of the squared-gradient
term, $K$, was evaluated using a recently-derived approximation based on the
potential\cite{Lutsko2011a}.

The most direct approach towards the determination of the nucleation pathway
based on the field equations, Eq.(\ref{sa}) and Eq.(\ref{s1}), would be to
discretize them by first discretizing the radial coordinate, $r\longmapsto
r_{i}\equiv\delta $, for some small length $\delta $, which then implies a
discretization of the density, $\rho \left( r\right) \longmapsto \rho
_{i} \equiv\rho \left( i\delta \right) $. Since the density is only expected to
vary quickly near the interface, the efficiency of this scheme can be
significantly improved by allowing for a dynamic discretization whereby the
density and position of the lattice points are allowed to vary. Here, this is
done by introducing a particular model for the density based on a
continuous, piecewise-linear approximation,%
\begin{equation}
\rho \left( r\right) =\left\{ 
\begin{array}{c}
\rho _{0},\;\;r<w_{0} \\ 
\rho _{0}+\left( \rho _{1}-\rho _{0}\right) \frac{r-w_{0}}{w_{1}}%
,\;\;w_{0}<r<w_{0}+w_{1} \\ 
\rho _{1}+\left( \rho _{2}-\rho _{1}\right) \frac{r-w_{0}-w_{1}}{w_{2}}%
,\;\;w_{0}+w_{1}<r<w_{0}+w_{1}+w_{2} \\ 
... \\ 
\rho _{\infty },\;\;w_{0}+...+w_{N-1}<r%
\end{array}%
\right.  \label{PW}
\end{equation}%
The parameters that are allowed to vary freely and dynamically are the
combinations $\rho _{i},w_{i}$ (referred to as "links") while the density $%
\rho _{\infty }$ is always fixed at that of the surrounding bulk (e.g. the
vapor density for nucleation of liquid droplets). Note that the simplest
model consisting of just two links, 
\begin{equation}\label{links}
\rho \left( r\right) =\left\{ 
\begin{array}{c}
\rho _{0},\;\;r<w_{0} \\ 
\rho _{0}+\left( \rho _{\infty }-\rho _{0}\right) \frac{r-w_{0}}{w_{1}}%
,\;\;w_{0}<r<w_{0}+w_{1} \\ 
\rho _{\infty },\;\;w_{0}+w_{1}<r%
\end{array}%
\right.
\end{equation}%
gives the minimal description of a cluster in terms of a radius, say $w_{0}$%
, an interfacial width, $w_{1}$, and a central density, $\rho _{0}$. Similar
parameterizations based on these three quantities are common in the
literature and here I note in particular one introduced by da Gama and Evans%
\cite{GamaEvans} and used recently by Ghosh and Ghosh\cite{Ghosh} to study both planar
interfaces and spherical clusters,%
\begin{equation}
\rho \left( r\right) =\left[ \rho _{0}-\frac{\rho _{0}-\rho _{\infty }}{2}%
\exp \left( a\left( r-R\right) \right) \right] \Theta \left( R-r\right) +%
\left[ \rho _{\infty }+\frac{\rho _{0}-\rho _{\infty }}{2}\exp \left(
-a\left( r-R\right) \right) \right] \Theta \left( r-R\right)  \label{exp}
\end{equation}%
where the radius is here denoted $R$. The parameters occurring in all of
these approximations are "order parameters"\ in the sense discussed above
and the most likely path of the density can be determined by means of the
order-parameter equations, Eq.(\ref{g1}). These are of course only
approximations but the piecewise-linear scheme becomes exact as the number
of links is increased.

As discussed above, once a parametrization is chosen, there are three steps
involved in determining the MLP. The first is to determine the critical
cluster, i.e. the saddle point of the free-energy surface in
parameter space. This was done using standard eigenvector-following methods%
\cite{Wales,Eigen} which result in a given set of parameters describing the
approximation to the critical cluster, $\mathbf{x}_{c}$. At the saddle
point, the Hessian matrix determined from the free energy has one negative
eigenvalue and a corresponding eigenvector, $\mathbf{v}_{\mathrm{neg}}$, which
specifies the unstable direction. The second and third parts of the
calculation are the  numerical integration of the order-parameter
equations, Eq.(\ref{g1}), with initial conditions consisting of a small
perturbation away from the critical point in the unstable direction, $%
\mathbf{x}_{c}\pm \epsilon \mathbf{v}_{\mathrm{neg}}$where the parameter $\epsilon $
is chosen so that the change in free energy is some specified amount. The
result of one of the perturbations will be that the system falls back to the
metastable state, i.e. the cluster will diminish, while the other will send
the system towards the final, stable state which means the cluster will grow.

\subsection{Nucleation of droplets}

The critical cluster for temperature $k_{B}T/\epsilon =0.375$ and
supersaturation $S\equiv \rho _{v}/\rho _{vc}=1.175$, where $\rho _{vc}$ is
the vapor density at coexistence, was determined for both the
piecewise-linear profiles with varying numbers of links and the exponential
profile. The "exact" critical cluster was determined by solving 
\begin{equation}
0=\frac{\delta \Omega \left[ \rho \right] }{\delta \rho \left( \mathbf{r}%
\right) }=K\nabla ^{2}\rho \left( \mathbf{r}\right) -f^{\prime }\left( \rho
\left( \mathbf{r}\right) \right) +\mu  \label{sga}
\end{equation}%
numerically using a relaxation technique\cite{NR} where the chemical
potential is simply $\mu =f^{\prime }\left( \rho _{v}\right) $. The excess
free energy and excess mass of the critical cluster is given in Table 1 as
calculated from several approximations. It demonstrates the gradual
convergence of the properties of the critical cluster to the exact result as
the number of links in the piecewise-linear profiles is increased. The
exponential profile gives a relatively good description of the critical
cluster based on only three parameters. Figure 1 shows the exact density
profile of the critical cluster and the piecewise profile based on 2, 4 and 10
links. It is apparent that while the simple profile based on 2 links is
quite crude, the 10-link profile is already a good approximation to the
continuous profile.

\begin{table}[tbp]
\caption{Excess free energy, $\Delta \Omega$,  and mass (number of molecules, $\Delta N$) for different
density models. The label ``PW$M$'' refers to piecewise-linear profiles with
$M$ links, Eq.(\ref{PW}).} \label{tab1}%
\begin{ruledtabular}
\begin{tabular}{ccc}
Approximation & $\beta \Delta \Omega $ & $\Delta N$ \\ \hline
PW2 & 98.9 & 1471 \\ 
PW4 & 82.9 & 1250 \\ 
PW6 & 79.4 & 1211 \\ 
PW8 & 77.8 & 1186 \\ 
PW10 & 77.1 & 1175 \\ 
Exponential, Eq.(\ref{exp}) & 80.4 & 1246 \\ 
Exact, Eq.(\ref{sga}) & 75.8 & 1158 \\
\end{tabular}
\end{ruledtabular}%\footnotetext[1]{using approximately 2000 atoms.}
\end{table}

\begin{figure}[tbp]
\includegraphics[angle=-90,scale=0.3]{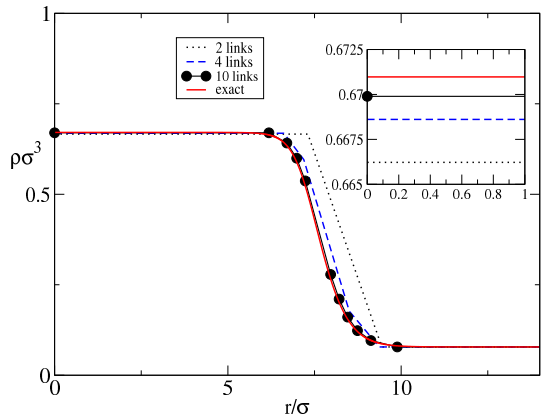} 
\caption{The density profile
of the critical cluster as determined by numerically finding a stationary
solution of Eq.(\ref{sga}), the ``exact'' solution, and from the piecewise
linear model with $N$ links, Eq.(\ref{PW}). The inset shows the convergence
of the central density towards the exact value as the number of links
increases. For the model with 10 links the figure shows the boundaries of
the links as circles. Notice how the density of the discretization is
automatically adjusted to be highest in the region of most rapid variation
of the profile.} \label{fig1}
\end{figure}

The evolution of the droplet was determined beginning with the profile for
the critical nucleus. The Hessian for the free energy was evaluated and
diagonalized giving in all cases a single negative eigenvalue. The profile
was perturbed by adding a multiple of the corresponding eigenvector with the
coefficient chosen so as to give a change in free energy of $1k_{\mathrm{B}}T$. The
equations for the MLP, expressed in terms of order parameters, Eq.(\ref{g1}%
), were then integrated using the Intel ODE solver library\cite{Intel}.
Depending on the sign of the coefficient used to perturb the state the
system would either fall backwards towards the metastable state (i.e.
decreasing cluster size) or forwards towards the stable state (increasing
cluster size). The integration was stopped in both cases when the excess
free energy reached $1k_{\mathrm{B}}T$. For the forward (growing) direction, this was
simply a convenient place to stop but in the backward direction, some such
cutoff is necessary since the weak-noise assumption must break down.

\begin{figure}[tbp]
\includegraphics[angle=-90,scale=0.3]{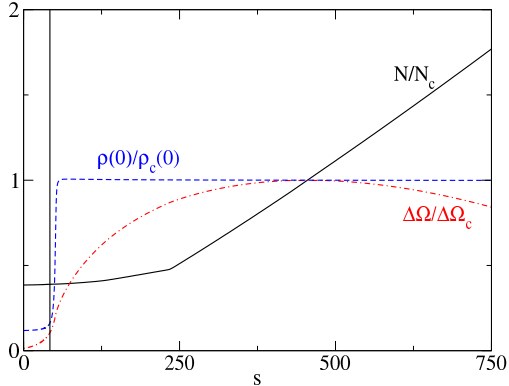} 
\caption{The variation with
path distance, $s$ as defined in Eq.(\ref{sdef}), of the excess number of
molecules, $N$, total excess free energy, $\Delta \Omega$, and central
density, $\rho(0)$, relative to the values for the critical cluster for a
nucleating droplet. The calculations were performed using the
piecewise-linear model with 10 links (19 parameters). For reference, the
values for the critical cluster are $N=1175$, $\Delta \beta \Omega = 77.1$ and
$\rho(0)\sigma^3 = 0.68$. To aid in comparison, the vertical line picks out
the same value of $s$ as in Fig.(\ref{fig3}).} \label{fig2}
\end{figure}

The independent variable in the integrations is time but care must be taken
in interpreting the results. The "time"\ has the physical meaning of the
time taken by the deterministic dynamics to drive the system in the desired
direction. When moving forwards in time, this is therefore the physical
time. However, when evaluating the MLP via the time-reversed dynamics, this
"time" does not correspond to the physical time needed for fluctuations to
drive the system up the free-energy gradient. Discussion as to what physical
meaning \emph{can} be attached to this time can be found in Bier et al\cite%
{Bier}. In order to avoid any ambiguity, the results here are displayed in
terms of the dimensionless distance along the path defined as 
\begin{equation}  \label{sdef}
s=\pm \sigma ^{-2}\int_{t_{1}}^{t_{2}}\sqrt{g_{ab}\frac{dx_{a}}{dt}%
\frac{dx_{b}}{dt}}dt
\end{equation}%
with $s=0$ being taken to correspond to the smallest cluster and the sign
being chosen so that $s$ increases monotonically moving from the smallest
cluster to the critical cluster and on to post-critical clusters.

\begin{figure}[tbp]
\includegraphics[angle=-90,scale=0.3]{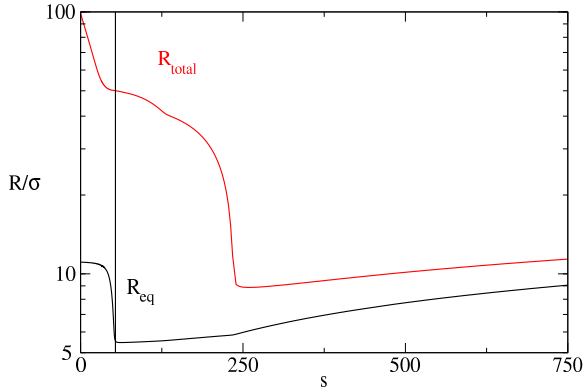} 
\caption{The variation of
the equimolar radius, $R_{eq}$, and the total radius, $R_{total}$, as
functions of distance along the path, $s$  as defined in Eq.(\ref{sdef}),
for a nucleating droplet. To aid in comparison, the vertical line picks out
the same value of $s$ as in Fig.(\ref{fig2}).} \label{fig3}
\end{figure}

Figure \ref{fig2} shows the evolution of the central density, the excess
mass and the excess free energy of the cluster as determined using the
piecewise-linear parametrization with 10 links (19 parameters). At first,
the central density remains very close to that of the vapor and the excess
free energy is small - on the order of a few $k_{B}T$. Surprisingly, even at
the very beginning of the process the excess mass is finite and a
significant fraction of the mass of the critical cluster. After some time
during which the density and mass increase very slowly the growth enters a
new regime in which the density increases rapidly to nearly the bulk liquid
density. From that point onward, the evolution is unremarkable as the
cluster grows to criticality and beyond. Corresponding behavior is seen in
the size of the cluster which is characterized in Fig. \ref{fig3} in two
ways: the equimolar radius, $R_{\mathrm{eq}}$ defined via $\frac{4\pi}{3}R_{\mathrm{eq}}^3(\rho(0)-\rho_{\infty})=\int \left(\rho(\mathbf{r})-\rho_{\infty}\right)d\mathbf{r}$,  and the total spatial extent, defined for the model given in Eq.(\ref{links}), as $R_{\mathrm{total}}\equiv\sum_{i=0}^{N-1}w_i$. (For a general representation of the density, this might be
characterized as the distance at which the density reaches some small
threshold above the background.) Initially, the equimolar radius is nearly
constant while the cluster has very large spatial extent and,
counter-intuitively, the cluster \emph{shrinks}. At the same point as the
density begins its rapid increase, the equimolar radius also begins to
increase even while the total spatial extent continues to diminish.
Eventually, the latter reaches a minimum and the cluster grows according to
both measures.

\begin{figure}[tbp]
\includegraphics[angle=-90,scale=0.3]{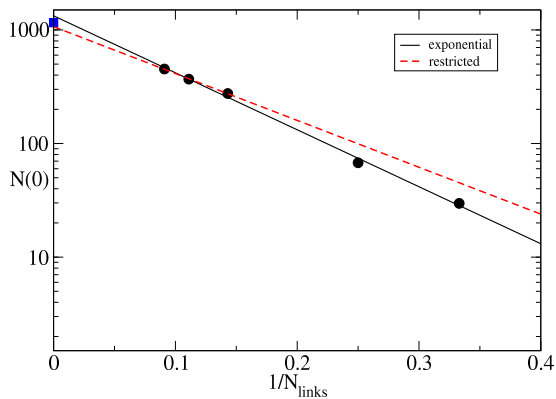} 
\caption{The excess number of
molecules in the cluster at the beginning of the nucleation process, $N(0)$,
as a function of the inverse of the number of links in the profile,
$N_{\mathrm{links}}$. The log-linear plot shows a nearly linear variation indicating
that $N(0) \sim Ae^{b/N_{\mathrm{links}}}$. Fitting all of the data to this functional
form and extrapolating gives an asymptotic value of $N(0)=1331$ and using
only the three points with the largest value of $N_{\mathrm{links}}$ gives
$N(0)=1068$. These values are consistent with the expectation that the
limiting value should be the mass of the critical cluster, $N_c=1175$, shown
as the square symbol.} \label{fig4}
\end{figure}

\begin{figure}[tbp]
\includegraphics[angle=-90,scale=0.3]{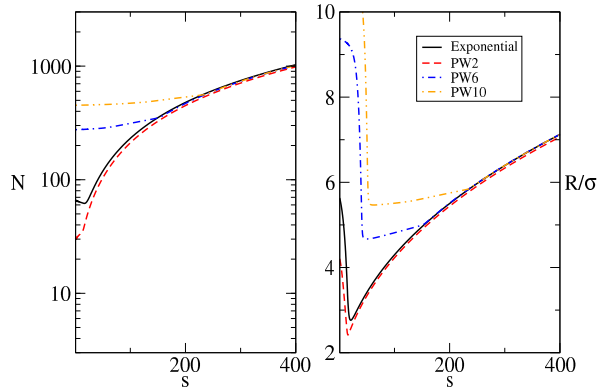} 
\caption{The excess number
of molecules, $N$, (left panel) and the total radius, $R$, (right panel) as
functions of distance along the nucleation pathway using piecewise-linear
models with $N$ links, ``PWN'', and the exponential profile. The figures
demonstrate that the short-time behavior is sensitive to how well the
density profile is represented whereas all of the models converge at longer
times. The exponential profile behaves very similarly to the PW2 profile,
which has the same number of free parameters, but gives a better estimate of
the properties of the critical cluster (see Table \ref{tab1}).} \label{fig5}
\end{figure}

These results are in stark contrast to the usual picture of droplet growth
as assumed in the Becker-D\"{o}ring picture (where droplets begin as dimers and
grow by attachment and detachment of monomers) as well as the corresponding
results from constrained DFT calculations (where droplets begin as zero mass
objects that grow monotonically). In fact, how can a droplet "begin" with a
finite mass?\ I shall attempt to answer this question from two different
perspectives: first, that of the formalism and second from a physical point
of view. In terms of the formalism, the MLP going from "nothing" to the critical
droplet is the time-reversed evolution of the dynamics with no noise:\ in
other words, it is the time-reversal of the deterministic evolution starting
with a droplet slightly smaller than the critical droplet. What happens in
such a case is obvious:\ the droplet evaporates by shedding mass to the
bulk. Since mass is conserved, the excess mass originally in the droplet
cannot just disappear but must diffuse over ever larger volumes until it is
lost in the (infinite) bulk. It is easy to understand, then, why the total
mass is almost constant in the beginning of the process. Indeed, one would
expect that the total mass should be the same as that of the critical
cluster and should not change at all. That this is the actual result, if one
could solve the model with the number of parameters tending to infinity, is
indicated in Fig. \ref{fig4} which shows the initial mass as a function of
the number of links in the profile. It is clear that the mass increases with
increasing refinement of the profile and that an extrapolation gives a mass
close to that of the critical cluster. The fact that the mass is less than
that of the critical cluster can be attributed to the difficulty of
approximating the density distribution when the density becomes very dilute.
Figure \ref{fig5} illustrates this as well, showing a large variation of the
various parameterizations at short times with close convergence at long
times. This includes the case of the exponential profile which is also not
well suited to representing the diffusive distribution at short times. Note
that according to this picture, one would expect the excess density at early
times to behave purely diffusively, giving, e.g., 
\begin{equation}
\Delta \rho \left( r,t\right) \sim \Delta N\left( 2\pi D\left\vert
t\right\vert \right) ^{-3/2}\exp \left( -\frac{r^{2}}{2D\left\vert
t\right\vert }\right)
\end{equation}%
(where "early times" means that $t$ is large and negative). Then, the total
mass would be conserved whereas the equimolar radius would be given by%
\begin{equation}
\Delta N=\frac{4\pi }{3}R_{eq}^{3}\left( t\right) \Delta \rho \left(
0,t\right) \Rightarrow R_{eq}\left( t\right) =\left( \frac{3}{4\pi }\right)
^{1/3}\left( 2\pi D\left\vert t\right\vert \right) ^{1/2}
\end{equation}%
thus explaining the slow change in the equimolar radius. The fact that the
result follows from such simple considerations suggests that, formally, it is
very robust being independent of the assumed free energy model,
potential or any other details.

From a physical perspective, the explanation given so far is not very
helpful. Instead, it is useful to divide the process into two parts as
indicated in Figs. \ref{fig2} and \ref{fig3}. During the first stage, matter
comes together "from infinity" in the form of a  diffuse structure. This
can be understood as the formation of a long-wavelength density fluctuation.
The presence of such fluctuations is expected and the fact that the
structure formed only costs a few $k_{B}T$ in energy supports this
expectation. What appears strange is the directed way that matter seems to
come together but this is simply a result of two artifacts. The first is the
assumption of spherical symmetry which means that the same thing must occur
in all directions at the same time; in reality, such a structure is
expected to be messy and ramified. Second is the nature of the MLP. It is
well known that the MLP generally follows the shortest possible path up the
gradient, with no backtracking or deviation\cite{Bier}. This gives it an
appearance of being directed towards reaching a final state (i.e. the
critical state) like a hiker deliberately climbing a hill. Any individual
realization of the stochastic process will, in contrast, consist of moves in
both directions - up and down the gradient - and will only, on average,
eventually trace out the MLP\ (or something like it). A final artifact is
that this is the MLP\ for a particular process: in other words, it
represents a conditional probability in which the condition is that one does
actually reach the critical state. A real system will make many abortive
attempts to climb the potential gradient and only after many such attempts
will it succeed. Here, in contrast, one studies the one successful attempt
which again gives the appearance of being directed.

So far, only the first part of the process - the formation of a density
fluctuation - has been discussed. The second part of the process - marked by
the rapid increase in density in the nascent cluster - can be interpreted as
the occurrence of an actual nucleation event. This part of the process
essentially proceeds classically and can be visualized in terms of the
Becker-D\"{o}ring picture or of the typical DFT\ calculations.

Putting these two pictures together gives a plausible and physically
appealing interpretation. Density fluctuations of all types occur in the
fluid and it makes intuitive sense that droplet nucleation should be more
common in regions of higher density since the free energy barrier will be
lower. Furthermore, the presence of a density fluctuation means that there
is excess mass from which to build the cluster. Even in the classical picture (which does not assume any pre-nucleation enrichment) a cluster must somehow draw
in material from the surrounding vapor but this part of the process is typically treated separately\cite{Lifshitz}, if at all. 
Considered in these terms, the classical process, with no enrichment, 
seems arguably more mysterious: a small, unstable cluster not only
survives but manages to draw in material and, in the fluid, mass flows at
such a rate that the cluster is always in contact with a vapor of exactly
the bulk density. In the present picture, the needed mass is already locally
present so that the formation of the cluster does not lead to the formation
of a depletion zone or require the orchestrated flow of matter from far
away. \ And finally, let us note that it is not the case that a fluctuation
with exactly the right density appears in order to host the nucleation
process: rather, fluctuations of all different sizes occur and if nucleation
begins in one that is too small, then it may fail to reach the critical
size. On the other hand, it is likely that the probability of a fluctuation
decreases with its size so that the \emph{most likely} scenario is of a
successful nucleation event occurring in a fluctuation which is just large
enough to supply the needed mass. This is the case that the MLP\ picks out.

\subsection{Nucleation of bubbles}

\begin{figure}[tbp]
\includegraphics[angle=-90,scale=0.3]{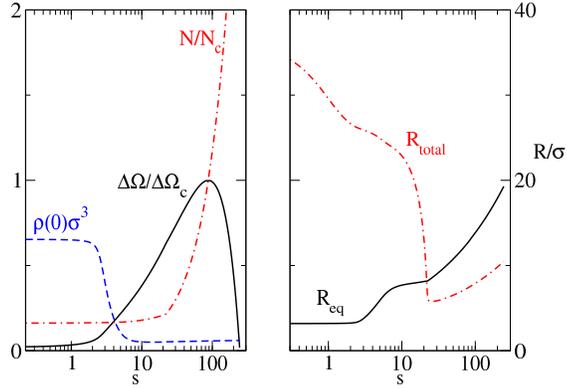} 
\caption{The left panel shows
the variation with path distance, $s$ as defined in Eq.(\ref{sdef}), of the
excess number of molecules, $N$, the  total excess free energy, $\Delta
\Omega$ and the central density, $\rho(0)$. The right panel shows the
evolution of the equimolar and total radii. The calculations were performed
using the piecewise-linear model with 8 links (15 parameters). The excess
number and free energies are scaled to the values for the critical cluster
which are  $N=-546$, $\Delta \Omega = 48.2k_BT$. } \label{fig6}
\end{figure}

As a second application, the evolution of the reverse process - the
formation of a low-concentration \textquotedblleft bubble\textquotedblright\
within the high-concentration \textquotedblleft liquid\textquotedblright\ -
was determined. For these calculations, the temperature was the same as
previously, $k_{B}T/\epsilon =0.375$ but the supersaturation was taken to be 
$S\equiv \rho _{v}/\rho _{vc}=0.9875$ so that the solution is
under-saturated making the dense phase unstable towards nucleation of the
less dense phase. The results are shown in Fig.(\ref{fig6}). A bubble
consists of a mass deficit relative to the bulk fluid and so,  according to the deterministic dynamics, a sub-critical
bubble is expected to vanish by
gradually spreading the deficit over ever larger volumes. Therefore, based
on the discussion of droplet nucleation, one would expect the process of
bubble nucleation to begin with the formation of a long-wavelength, small
amplitude deficit of density within which the bubble would nucleate. The
results of the numerical calculation bear this out and all of the preceding
discussion of droplet nucleation can be carried over to the dual process of
bubble nucleation.

\subsection{Order parameter dynamics}

As a final illustration of the formalism, the general expressions for the
order-parameter dynamics, Eq.(\ref{final}), are specialized to give a
CNT-level of description of the nucleation process for either droplet or
bubble nucleation. By "CNT-level" it is meant that there will be only a
single order parameter, the radius of the cluster $R$, and the calculations
will be performed to lowest order in $w/R$ where $w$ is the width of the
interface. The first step is the evaluation of the metric. This is most
easily done for a concrete model and here the minimal piece-wise linear
model with 2 links will be used. The central density is fixed at the bulk
density of the new phase, $\rho _{0}$, and the width $w$ is held constant.
Then, a straightforward calculation gives%
\begin{equation}
g_{RR}=\int_{0}^{\infty }\frac{1}{4\pi r^{2}\rho \left( r\right) }\left( 
\frac{\partial m\left( r\right) }{\partial R}\right) ^{2}dr=\frac{4\pi
\left( \rho _{0}-\rho _{\infty }\right) ^{2}}{\rho _{\infty }}R^{3}\left(
1+O\left( w/R\right) \right)
\end{equation}%
where $\rho _{\infty }$ is the density of the initial metastable phase.
Using this, the SDE for the order parameter becomes%
\begin{equation}
\frac{dR}{dt}=-D\frac{\rho _{\infty }}{4\pi \left( \rho _{0}-\rho _{\infty
}\right) ^{2}R^{3}}\frac{\partial \beta \Omega }{\partial R}+\sqrt{2D\frac{%
\rho _{\infty }}{4\pi \left( \rho _{0}-\rho _{\infty }\right) ^{2}R^{3}}}\xi
\left( t\right)
\end{equation}%
or, written in terms of the mass of a cluster, $N=\frac{4\pi }{3}R^{3}\rho
_{0}$,%
\begin{equation}
\frac{dN}{dt}=-D\frac{4\pi \rho _{0}^{2}\rho _{\infty }}{\left( \rho
_{0}-\rho _{\infty }\right) ^{2}}R\left( N\right) \frac{\partial \beta
\Omega }{\partial N}+\sqrt{2D\frac{4\pi \rho _{0}^{2}\rho _{\infty }}{\left(
\rho _{0}-\rho _{\infty }\right) ^{2}}R\left( N\right) }\xi \left( t\right)
\end{equation}%
Recalling that this equation must be interpreted in the Stratonovich sense,
it is equivalent to a Fokker-Planck equation for the probability of observing
a cluster of size $N$ at time $t$, $p\left( N,t\right) $, of the form\cite%
{Gardiner}%
\begin{equation}
\frac{\partial p\left( N,t\right) }{\partial t}=\frac{\partial }{\partial N}%
\left[ D\frac{4\pi \rho _{0}^{2}\rho _{\infty }}{\left( \rho _{0}-\rho
_{\infty }\right) ^{2}}R\left( N\right) \left( \frac{\partial \beta \Omega }{%
\partial N}+\frac{1}{6N}\right) +D\frac{4\pi \rho _{0}^{2}\rho _{\infty }}{%
\left( \rho _{0}-\rho _{\infty }\right) ^{2}}R\left( N\right) \frac{\partial 
}{\partial N}\right] p\left( N,t\right)  \label{fp1}
\end{equation}%
This can be compared to the classical result of Zeldovich (see Ref. %
\onlinecite{Kashchiev}, Eq. (9.27)) which is derived using the Becker-D%
\"{o}ring picture,%
\begin{equation}
\frac{\partial p\left( N,t\right) }{\partial t}=\frac{\partial }{\partial N}%
\left[ f\left( N\right) \frac{\partial \beta \Omega }{\partial N}+f\left(
N\right) \frac{\partial }{\partial N}\right] p\left( N,t\right)
\end{equation}%
where the monomer attachment frequency, $f(N)$, in the case of
diffusion-limited kinetics is given by\ (see Ref.\onlinecite{Kashchiev}, Eq.
10.18) 
\begin{equation}
f\left( N\right) =\eta 4\pi R\left( N\right) D\rho _{\infty }
\end{equation}%
and where all unknown details of the monomer attachment process are
contained in the sticking coefficient, $\eta $. In the CNT limit of large $%
N$, the factor of $1/(6N)$ in Eq.(\ref{fp1}) can be ignored and it then agrees
with the Zeldovich result with the sticking coefficient identified as%
\begin{equation}
\eta =\frac{\rho _{0}^{2}}{\left( \rho _{0}-\rho _{\infty }\right) ^{2}}
\end{equation}%
For the nucleation of droplets from vapor, one expects that $\rho _{0}\gg\rho
_{\infty }$ so that this gives $\eta \approx 1$. This shows that by
specializing the general theory to the CNT regime of large clusters and a
single order parameter, one is able to recover all of the elements of the
classical formalism.

Similarly, the free energy in the squared-gradient approximation is\cite{Lutsko2011a}%
\begin{equation}
\Omega \left[ \rho \right] -\Omega _{\infty }=4\pi \frac{R^{3}}{3}\left(
\omega _{0}-\omega _{\infty }\right) +4\pi R^{2}\gamma
\end{equation}%
where the excess surface free energy is 
\begin{equation}
\gamma =w\overline{\omega }+K\frac{\left( \rho _{\infty }-\rho _{0}\right)
^{2}}{2w}\left( 1+O\left( \frac{w}{R}\right) \right)
\end{equation}%
and the density-averaged free energy per unit volume is%
\begin{equation}
\overline{\omega }=\frac{1}{\left( \rho _{0}-\rho _{\infty }\right) }%
\int_{\rho _{\infty }}^{\rho _{0}}\left( \omega \left( x\right) -\omega
_{\infty }\right) dx
\end{equation}%
For a large interface, the width can be estimated by minimizing the free
energy of the critical cluster giving%
\begin{equation}
w=\sqrt{\frac{\left( \rho _{0}-\rho _{\infty }\right) ^{2}K}{2\left( 
\overline{\omega }-\omega _{\mathrm{coex}}\right) }}
\end{equation}%
where $\omega _{\mathrm{coex}}$ is the free energy per unit volume at coexistence\cite%
{Lutsko2011a}. All quantities here are therefore directly determined by the
interaction potential.

Substituting these elements into Eq. (\ref{final}) gives%
\begin{equation}
\frac{dR}{dt}=-D\frac{\rho _{\infty }\left( \omega _{0}-\omega _{\infty
}\right) }{\left( \rho _{0}-\rho _{\infty }\right) ^{2}}R^{-1}\left( 1-\frac{%
R_{c}}{R}\right) -\sqrt{D\frac{\rho _{\infty }\left( \omega _{\infty}-\omega
_{0 }\right) }{2\pi \left( \rho _{0}-\rho _{\infty }\right) ^{2}}R^{-3}}\xi
\left( t\right)
\end{equation}%
where%
\begin{equation}
R_{c}=\frac{2\gamma }{ \omega _{\infty}-\omega _{0 }}
\end{equation}%
is the usual CNT expression for the critical radius. Note that for a single
order parameter, the anomalous force vanishes and so this model applies to
the strong-noise regime. In principle, it can be used to determine a mean
first-passage time and, hence, nucleation rate as will be discussed elsewhere. For now, it is interesting to make one further observation
which is that for large, \emph{post-critical} clusters, where the noise
becomes unimportant, this gives the deterministic growth law%
\begin{equation}
\frac{dR}{dt}\approx -D\frac{\rho _{\infty }\left( \omega _{0}-\omega
_{\infty }\right) }{\left( \rho _{0}-\rho _{\infty }\right) ^{2}}R^{-1}
\end{equation}%
implying that $R\sim t^{1/2}$ which is the well-known result for
diffusion-limited cluster growth\cite{Saito,[{}][{; see ch. 12, first
problem}]Lifshitz}.

\section{Conclusions}

The goal of this work is a description of nucleation based on a
dynamical, non-equilibrium description of fluctuations with the additional
aim of making contact with the usual ideas of classical nucleation theory.
The purpose is to develop a consistent and unique description
of nucleation pathways, kinetics and post-critical cluster growth. Starting
with fluctuating hydrodynamics this goal is achieved for the particular
case of Brownian particles where the limit of strong dissipation provides
several simplifications. The nucleation pathway is characterized as the the
most likely path and a reduced description in terms of order parameters is
developed. The order parameter description includes both a Langevin dynamics
for the order parameters and a description of the most likely path. It is
noted that in the weak noise limit, all levels of description lead to a
unique description of the MLP that passes through the saddle point of the
free energy functional and that is determined by steepest-descent on the
free energy surface governed by a metric that is unambiguously specified. At
this level, the goal of making contact with CNT is achieved with the
role of the free energy emerging from the nonequilibrium description rather
than being assumed \textit{a priori}. Furthermore, as illustrated in the simple case
of a single order parameter, one also has a Langevin description of the
process which can be used to develop a rate theory that recovers the
classical results and is consistent with the expected post-critical growth
law. Thus, the goal of giving a unified description of these three elements
- rates, pathways and growth-laws - is also achieved.

One of the main theoretical results concerns the determination of the
nucleation pathway as characterized by the Most Likely Path. It is shown
that, in the limit of weak noise, this is determined by the deterministic
part of the dynamics moving either forward or backward in time away from the
critical point. This result represents a generalization of
Dynamic Density Functional Theory to problems involving barrier crossing.
The same result holds when the dynamics are written in terms of order
parameters. In all cases, the MLP can, alternatively, be viewed as being
determined by steepest-descent on the free energy surface under a prescribed
metric. It is noted that the strong-noise limit may also be handled but
investigation of the effect of strong noise is deferred to a later date. In
general, it is seen that one recovers the elements of CNT in the weak-noise
limit which therefore plays a role analogous to the quasi-classical limit in
quantum theory.

It is interesting to compare this result with previous approaches. As
discussed in the Introduction, there have been two primary approaches to the
description of nucleation pathways using DFT. The first is the minimization
of the free energy under a constraint that stabilizes pre-critical and
post-critical clusters. There is no direct analogy to this method in the
present theory. The other approach that has more recently begun to be used
is steepest-descent on the free energy surface. The problem then  is that it
is necessary to define a distance in the parameter space (whether it be the
space of density distributions or a space of parameters) and there has not
been a compelling, much less unique, prescription for doing so. In Refs. [%
\onlinecite{LutskoBubble1, LutskoBubble2}] it was proposed that  a natural
choice was the Euclidean distance in density space, 
\begin{equation}
d^2[\rho_1,\rho_2] = \int (\rho_{1}(\mathbf{r})-\rho_{2}(\mathbf{r}))^2d%
\mathbf{r}
\end{equation}
giving, for arbitrary parameters, 
\begin{equation}
g^{\rho}_{ab}=\int \frac{\partial \rho(\mathbf{r})}{\partial x_a}\frac{%
\partial \rho(\mathbf{r})}{\partial x_b} d\mathbf{r}
\end{equation}
which should be contrasted with the present results, Eq.(\ref{genmetric})
and Eq.(\ref{metric}). The differences between these metrics are not
trivial: integration of the steepest-descent equations, Eq.(\ref{g1}), using
the heuristic $g^{\rho}_{ab}$ gives completely different results for droplet
nucleation from those presented above. In fact, the result is the
``classical'' one whereby a droplet  begins as a cluster of zero mass and
the mass slowly increases as the density and radius increases during cluster
formation\cite{LutskoBubble2, Lutsko2011a}. It is interesting to note that
the difference in the results is not due to the factor of the inverse
density in the expression for the correct metric, but rather is due to the
fact that it is based on the cumulative mass distribution rather than the
density. In retrospect, this could be argued to be quite physical: in moving
from one density configuration to another, what actually is required is the
movement of mass from one place to another so that the ``closeness'' of one
configuration to another is more naturally characterized by how different
they are in mass distribution. Furthermore, it is to be expected to result
from a formulation that respects the conservation of mass and in which,
therefore, the rate of transport of mass is a limiting factor. Conversely,
the incorrect, heuristic metric  would result from determining the MLP from a dynamics of the form 
\begin{equation}
\frac{d \rho(\mathbf{r},t)}{dt}= -D\frac{\delta \Omega[\rho]}{\delta \rho(\mathbf{r%
},t)} + \sqrt{2Dk_BT}\xi(\mathbf{r},t)
\end{equation}
which is to say, a non-conserved dynamics. (In this equation, $\Omega[\rho] = F[\rho] - \mu N[\rho]$ for fixed chemical potential, $\mu$. In other words, this equation for a non-conserved total mass is to be understood in the grand-canonical ensemble.) While this could be appropriate
for describing some types of simulations that do not conserve mass, it is
clearly inappropriate for real physical systems. Finally, it goes without
saying that doing steepest-descent using an ad-hoc metric based on some
parametrization of the free energy is the same thing as assuming an ad hoc
dynamics which will not, in general, be physically relevant.

Another point that has not been emphasized is the ability of this formalism
to distinguish between multiple pathways. Just as there may be multiple
local minima, and only one true minimum, of a function, there can be
multiple local MLP's and only one truly \emph{Most} Likely Path. The
simplest and most relevant case occurs when the free energy surface
possesses multiple saddle points. For example, in the case of protein
nucleation there is a question of whether the system proceeds directly from
low-concentration solution to solid or whether it proceeds in two steps by
passing through an intermediate, metastable high-concentration phase of the
type studied above. Until now, the only question that was accessible
theoretically was the height of the energy barriers and this alone was
assumed to determine the relevant path. However, this ignores the role of
dynamics (for a very interesting discussion of this point see Ref. %
\onlinecite{Whitelam}). The present theory allows a way to address this
problem. First, it is possible to determine local MLP's e.g., in the
weak-noise limit by means of steepest descent from the various saddle
points. Then, given these paths, their relative probabilities are easily
determined by evaluating Eq.(\ref{Graham1})-(\ref{Graham3}). The absolute
probability of the paths requires knowledge of possibly unknown
normalization constants but the relative probabilities can be determined
directly by taking the ratio the results. This will be elaborated at a later
time.

When applied to globular proteins, namely the case of nucleation of the high-concentration amorphous
phase from the low-concentration solution, an
unexpected result is found whereby the process involves two steps. In the
first step, a long-wavelength, low-amplitude density fluctuation occurs. In
the second step, a nucleation event takes place within this region of
enhanced density. I argued that this is to be expected since, on
the one hand, such density fluctuations \emph{do} occur, and at little cost
in free energy, and, on the other hand, nucleation should be easier in such
a region since the free energy barriers should be lower and, equally
importantly, since the excess mass required for cluster formation is
present. This result is very robust in that it is independent of the
interaction potential and of the free energy-functional used while  it is valid in the weak-noise limit and could be modified when the effects of strong noise
are considered. It is interesting that some of the
free-energy-with-constraint methods have given hints of a similar pathway,
most notably as discussed in Ref. 
\onlinecite{Lutsko2011a,
EvansArcherNucleation}. It is unclear whether this result is of any
practical significance although one might speculate that the enhancement or
suppression of long-wavelength density (concentration) fluctuations could
have a strong effect on the nucleation rate.

The theory developed here is specific to the case of Brownian particles.
In some sense, however, it can rather be viewed more as an illustration of how a
theory can be constructed in other cases. For example, for pure fluids, one
still has fluctuating hydrodynamics and the basic ideas can be developed in
a similar way. In that case, temperature fluctuations and energy transport
are important so that the fluctuating hydrodynamic theory will have to be extended
to include the heat equation. The main complication might be that the
hydrodynamic theory cannot be collapsed into a single equation for the density,
but rather that one must deal will multiple quantities throughout. This may
limit the practical utility of the theory but it could still yield useful
insights into, e.g., the MLP in the weak noise limit and the role of the
free energy.

Future directions for further work include the investigation of the role of
strong noise. Also interesting would be the application to heterogeneous
nucleation, nucleation in confined systems and the nucleation of solids.

\begin{acknowledgments}
It is a pleasure to thank Gr\'{e}goire Nicolis for several insightful
comments. I am also grateful to Nelido Gonz\'{a}lez-Segredo for extensive comments on an early draft of this paper. This work was supported in part by the European Space Agency under
contract number ESA AO-2004-070.
\end{acknowledgments}

\appendix{}

\section{Ito and Stratonovich interpretations are the same\label%
{Interpretation}}

In general, using the Riemann-Stieltjes definition of integration, stochastic
integrals differ from non-stochastic integrals in that their value depends
on where the sampling point is chosen within each interval. This means that
stochastic differential equations differ according to whether they are
interpreted in terms of Ito or Stratonovich interpretations. However, in
ordinary fluctuating hydrodynamics, the two interpretations turn out to be
the same\cite{Saarloos}. Here, it is shown that this is also true for the
over-damped dynamical model in both the general and spherically symmetric cases. Note that this does \emph{not} hold for the approximate order-parameter dynamics thus giving rise to the anomalous force discussed in Section \ref{opd} and in Appendix \ref{Noise}. Incidentally, it is worth noting that the fact that the two interpretations differ in the latter case is not directly attributable to any approximation: Ito-Stratonovich equivalence can be broken simply by a change in variables.

\subsection{Equivalence for the general model}

In general, the Stratonovich SDE 
\begin{equation}
\frac{dx_{i}}{dt}=F_{i}\left( \mathbf{a}\right) +M_{ij}\left( \mathbf{a}%
\right) \xi _{j}\left( t\right) ,\;\left\langle \xi _{j}\left( t\right) \xi
_{l}\left( t^{\prime }\right) \right\rangle =\delta _{jl}\delta \left(
t-t^{\prime }\right) 
\end{equation}%
corresponds to an Ito SDE of the form%
\begin{equation}
\frac{dx_{i}}{dt}=F_{i}\left( \mathbf{a}\right) +\frac{\partial M_{ij}\left( 
\mathbf{a}\right) }{\partial a_{k}}M_{kj}\left( \mathbf{a}\right)
+M_{ij}\left( \mathbf{a}\right) \xi _{j}\left( t\right) ,\;\left\langle \xi
_{j}\left( t\right) \xi _{l}\left( t^{\prime }\right) \right\rangle =\delta
_{jl}\delta \left( t-t^{\prime }\right) 
\end{equation}%
so the question comes down to an investigation of the second term on the
right in the Ito SDE, the so-called spurious drift. To evaluate this for the
hydrodynamic model, the procedure of van Saarloos\cite{Saarloos} et al is
used whereby the stochastic differential equation is discretized in the
spatial variable. Specifically, $\mathbf{r}\longmapsto \mathbf{r}_{\mathbf{n}%
}=\mathbf{n}\Delta $ where $\mathbf{n}=(n_{1},n_{2},...,n_{D})$ are the
(integer) coordinates of a lattice point in $D$ dimensions and $\Delta $ is
the lattice spacing. Also writing $\rho _{\mathbf{n}}\equiv \rho (\mathbf{r}%
_{\mathbf{n}})$, the noise term for the  over-damped model becomes%
\begin{equation}
\left[ \mathbf{\nabla \cdot }\sqrt{\rho \left( \mathbf{r}\right) }\mathbf{%
\xi }\left( \mathbf{r},t\right) \right] _{\mathbf{r}_{\mathbf{n}%
}}\rightarrow \frac{1}{2\Delta }\sum_{a=1}^{D}\left( \sqrt{\rho _{\mathbf{n}+%
\widehat{\mathbf{e}}^{(a)}}}\mathbf{\xi }_{a,\mathbf{n}+\widehat{\mathbf{e}}%
^{(a)}}-\sqrt{\rho _{\mathbf{n}-\widehat{\mathbf{e}}^{(a)}}}\mathbf{\xi }_{a,%
\mathbf{n}-\widehat{\mathbf{e}}^{(a)}}\right) 
\end{equation}%
where $\widehat{\mathbf{e}}^{(a)}$ is the unit vector in the $a$-direction
(i.e. $\widehat{\mathbf{e}}_{i}^{(a)}=\delta _{ia}$) and the time-arguments
have been suppressed. Introducing the notation $\delta _{\mathbf{n,n}%
^{\prime }}$ meaning that all components of the vectors $\mathbf{n}$ and $%
\mathbf{n^{\prime }}$ are the same the equivalent of $M_{ij}$ can be
identified as 
\begin{eqnarray}
M_{\mathbf{n},\mathbf{n^{\prime }}a} &=&\frac{1}{2\Delta }\left( \sqrt{\rho
_{\mathbf{n}+\widehat{\mathbf{e}}^{(a)}}}\delta _{\mathbf{n+\widehat{\mathbf{%
e}}^{(a)},n}^{\prime }}-\sqrt{\rho _{\mathbf{n}-\widehat{\mathbf{e}}^{(a)}}}%
\delta _{\mathbf{n-\widehat{\mathbf{e}}^{(a)},n}^{\prime }}\right)  \\
&=&\frac{1}{2\Delta }\sqrt{\rho _{\mathbf{n}^{\prime }}}\left( \delta _{%
\mathbf{n+\widehat{\mathbf{e}}^{(a)},n}^{\prime }}-\delta _{\mathbf{n-%
\widehat{\mathbf{e}}^{(a)},n}^{\prime }}\right) .  \notag
\end{eqnarray}%
We therefore need%
\begin{eqnarray}
&&\sum_{\mathbf{kn^{\prime }}a}\frac{\partial M_{\mathbf{n},\mathbf{n^{\prime }%
}a}}{\partial \rho _{\mathbf{k}}}M_{\mathbf{k},\mathbf{n^{\prime }}a} \\
&=&%
\frac{1}{8\Delta ^{2}}\sum_{\mathbf{kn}^{\prime }a}\frac{1}{\sqrt{\rho _{%
\mathbf{k}}}}\delta _{\mathbf{n}^{\prime },\mathbf{k}}\left( \delta _{%
\mathbf{n+\widehat{\mathbf{e}}^{(a)},n}^{\prime }}-\delta _{\mathbf{n-%
\widehat{\mathbf{e}}^{(a)},n}^{\prime }}\right) \sqrt{\rho _{\mathbf{n}%
^{\prime }}}\left( \delta _{\mathbf{k+\widehat{\mathbf{e}}^{(a)},n}^{\prime
}}-\delta _{\mathbf{k-\widehat{\mathbf{e}}^{(a)},n}^{\prime }}\right)  \notag \\
&=&\frac{1}{8\Delta ^{2}}\sum_{\mathbf{k}a}\left( \delta _{\mathbf{n+%
\widehat{\mathbf{e}}^{(a)},k}}-\delta _{\mathbf{n-\widehat{\mathbf{e}}%
^{(a)},k}}\right) \left( \delta _{\mathbf{k+\widehat{\mathbf{e}}^{(a)},k}%
}-\delta _{\mathbf{k-\widehat{\mathbf{e}}^{(a)},k}}\right)   \notag \\
&=&0  \notag
\end{eqnarray}%
thus proving equivalence of the two interpretations.

\subsection{Equivalence for the spherically symmetric model}

In the case of spherical symmetry one has that%
\begin{equation}
\frac{dm\left( r\right) }{dt}=4\pi r^{2}D\rho \left( r\right) \frac{\partial 
}{\partial r}\frac{1}{4 \pi r^2}\frac{\delta \beta F }{\delta \rho \left( r\right) }+\sqrt{%
D4\pi r^{2}\rho \left( r\right) }\xi \left( r\right)
\end{equation}%
We discretize as in the previous case but now only require a one-dimensional lattice. Writing $r \longmapsto r_{n}=n\Delta$ and 
 keeping in mind that 
\begin{equation}
4\pi r^{2}\rho \left( r\right) =\frac{\partial m\left( r\right) }{\partial r}
\end{equation}%
one finds the noise term
\begin{equation}
\sqrt{D\frac{m_{n+1}-m_{n-1}}{2\Delta }}\xi_n
\end{equation}%
where $m_n=m(r_n;t)$, etc.  Thus
\begin{equation}
M_{ij}=\sqrt{D\frac{m_{i+1} -m_{i-1} }{2\Delta }}\delta _{ij}
\end{equation}%
and so%
\begin{equation}
\frac{\partial M_{ij}\left( \mathbf{a}\right) }{\partial a_{k}}M_{kj}\left( 
\mathbf{a}\right) =\delta _{ik}\frac{\partial \sqrt{D\frac{%
m_{i+1} -m_{i-1} }{2\Delta }}}{\partial m_{k}}%
\sqrt{D\frac{m_{k+1} -m_{k-1} }{%
2\Delta }}=0
\end{equation}

\section{MLP for potential-driven dynamics with fluctuation-dissipation
relation}

\label{proof}

The purpose of this Appendix is to sketch a straightforward extension of the
results of Vanden-Eijnden and Heymann\cite{Heymann} whereby the assumption
of a constant Onsager matrix and white noise is lifted while still assuming
a fluctuation-dissipation relation.

\bigskip In the following, I consider a set of $N$ stochastic variables, $%
x_{i}\left( t\right) $, governed by a diffusive, gradient-driven stochastic
dynamics with multiplicative noise,%
\begin{equation}
\frac{d\mathbf{x}}{dt}=-\mathbf{L}\left( \mathbf{x}\right) \cdot \frac{%
\partial}{\partial\mathbf{x}}V\left( \mathbf{x}\right) +\sqrt {2\epsilon}%
\mathbf{\sigma}\left( \mathbf{x}\right) \cdot\mathbf{\xi}\left( t\right)
\label{D}
\end{equation}
where $\mathbf{L}\left( \mathbf{x}\right) $ is a state-dependent matrix of
kinetic coefficients, where the scalar constant $\epsilon$ and matrix $%
\mathbf{\sigma}\left( \mathbf{x}\right) $ determine the noise amplitude and
the noise itself is Gaussian, white and diagonally correlated%
\begin{equation}
\left\langle \xi_{i}\left( t\right) \xi_{j}\left( t^{\prime}\right)
\right\rangle =\delta_{ij}\delta\left( t-t^{\prime}\right)
\end{equation}
Note that the probability density for $\mathbf{\xi}\left( t\right) $ to
assume some value, say $\mathbf{z}$, is 
\begin{equation}
P\left( \mathbf{\xi}\left( t\right) =\mathbf{z}\right) =\left( \frac {1}{2\pi%
}\right) ^{N/2}\exp\left( -z^{2}/2\right)
\end{equation}
The key assumption in the following is that a fluctuation-dissipation
relation holds, namely%
\begin{equation}
\mathbf{L}\left( \mathbf{x}\right) =\mathbf{\sigma}\left( \mathbf{x}\right)
\cdot\mathbf{\sigma}^{\mathrm{T}}\left( \mathbf{x}\right) \equiv \mathbf{D}\left( 
\mathbf{x}\right)  \label{FDT}
\end{equation}
where the second equality reminds that the middle quantity is the diffusion
matrix occurring in the Fokker-Planck equation.

Following Refs.\onlinecite{Heymann1,Heymann2} the probability density for a
given path taking place from time $t=0$ to $t=T$ is $P=\exp\left( -\frac{1}{%
4\epsilon}S_{T}\left[ x\right] \right) $, where the action is 
\begin{equation}
S_{T}\left[ x\right] = \int_{0}^{T}\left( \frac{d\mathbf{x}}{dt}+\mathbf{L}%
\left( \mathbf{x}\right) \cdot\frac{\partial}{\partial \mathbf{x}}V\left( 
\mathbf{x}\right) \right) \cdot\mathbf{D}^{-1}\left( \mathbf{x}\right)
\cdot\left( \frac{d\mathbf{x}}{dt}+\mathbf{L}\left( \mathbf{x}\right) \cdot%
\frac{\partial}{\partial\mathbf{x}}V\left( \mathbf{x}\right) \right) dt
\end{equation}
and it must be remembered that $\mathbf{x}$ depends on time in this and all
following expression. A path between two points $\mathbf{x}_{1}$ and $%
\mathbf{x}_{2}$ is therefore a curve $\mathbf{x}\left( t\right) $ such that $%
\mathbf{x}\left( 0\right) =\mathbf{x}_{1}$ and $\mathbf{x}\left( T\right) =%
\mathbf{x}_{2}$. The most likely path (MLP) is determined by minimizing $%
S_{T}\left[ x\right] $ over both $\mathbf{x}\left( t\right) $, subject to
the constraints on the end points, and time, $T$.

In the following, we specialize to the situation that $\mathbf{x}_{1}$ and $%
\mathbf{x}_{2}$ are metastable points and in fact attractors.\ We assume
that x-space is divided by a separatrix into two regions:\ region I\ in
which points are attracted to $\mathbf{x}_{1}$ and region II\ in which
points are attracted to $\mathbf{x}_{2}$. The separatrix is a curve which
will be called $\mathcal{S}$. Any path from $\mathbf{x}_{1}$ to $\mathbf{x}%
_{2}$ must cross $\mathcal{S}$ at least once. For the moment, it will be
assumed that only one such crossing occurs and the possibility of multiple
crossings will be discussed below. Any such path can therefore be separated
into two pieces:\ one running from $\mathbf{x}_{1}$ to some point $\mathbf{x}%
_{s}\in\mathcal{S}$ and a second part from $\mathbf{x}_{s}$ to $\mathbf{x}%
_{2}$: 
\begin{align}
S_{T}\left[ x\right] & =S^{I}\left[ x\right] +S^{II}\left[ x\right] \\
S^{I}\left[ x\right] & =\int_{0}^{T_{s}}\left( \frac {d\mathbf{x}}{dt}+%
\mathbf{L}\left( \mathbf{x}\right) \cdot\frac{\partial }{\partial\mathbf{x}}%
V\left( \mathbf{x}\right) \right) \cdot\mathbf{D}^{-1}\left( \mathbf{x}%
\right) \cdot\left( \frac{d\mathbf{x}}{dt}+\mathbf{L}\left( \mathbf{x}%
\right) \cdot\frac{\partial}{\partial\mathbf{x}}V\left( \mathbf{x}\right)
\right) dt  \notag \\
S^{II}\left[ x\right] & = \int_{T_{s}}^{T}\left( \frac {d\mathbf{x}}{dt}+%
\mathbf{L}\left( \mathbf{x}\right) \cdot\frac{\partial }{\partial\mathbf{x}}%
V\left( \mathbf{x}\right) \right) \cdot\mathbf{D}^{-1}\left( \mathbf{x}%
\right) \cdot\left( \frac{d\mathbf{x}}{dt}+\mathbf{L}\left( \mathbf{x}%
\right) \cdot\frac{\partial}{\partial\mathbf{x}}V\left( \mathbf{x}\right)
\right) dt  \notag
\end{align}
with $\mathbf{x}\left( T_{s}\right) =\mathbf{x}_{s}$. Clearly, the MLP is
determined by minimizing over $\mathbf{x}_{s}\in\mathcal{S}$ and $0\leq
T_{s}\leq T$ as well. Consider the second term first. Once the
separatrix is crossed all points are attracted to $\mathbf{x}_{2}$ by hypothesis, so
the path%
\begin{align}
\frac{d\mathbf{x}}{dt} & =-\mathbf{L}\left( \mathbf{x}\right) \cdot \frac{%
\partial}{\partial\mathbf{x}}V\left( \mathbf{x}\right) \\
\mathbf{x}\left( T_{s}\right) & =\mathbf{x}_{s}  \notag
\end{align}
will eventually reach $\mathbf{x}_{2}$:\ the time required determines $T$
given $T_{s}$. (Note that one does not really start on the
separatrix but rather at a point infinitesimally near it on the region II
side.) However, this path has the property that $S^{II}\left[ x\right] =0$
and this is minimal since the integrand is positive definite. This is just
the trivial result that the deterministic path is the MLP\ if it passes
through the desired points.

It is not possible to take the deterministic path in region I since the
system must go from $\mathbf{x}_{1}$ to $\mathbf{x}_{s}$ and the
deterministic dynamics is assumed to always take points in region I towards $%
\mathbf{x}_{1}$. So, upon noting that expanding the action gives%
\begin{align}
S^{I}\left[ x\right] & = \int_{0}^{T_{s}}\left( \frac {d\mathbf{x}}{dt}\cdot%
\mathbf{D}^{-1}\left( \mathbf{x}\right) \cdot \frac{d\mathbf{x}}{dt}+\left( 
\mathbf{L}\left( \mathbf{x}\right) \cdot \frac{\partial}{\partial\mathbf{x}}%
V\left( \mathbf{x}\right) \right) \cdot\mathbf{D}^{-1}\left( \mathbf{x}%
\right) \cdot\left( \mathbf{L}\left( \mathbf{x}\right) \cdot\frac{\partial}{%
\partial\mathbf{x}}V\left( \mathbf{x}\right) \right) \right) dt \\
& +2\int_{0}^{T_{s}}\frac{d\mathbf{x}}{dt}\cdot\mathbf{D}^{-1}\left( \mathbf{%
x}\right) \cdot\mathbf{L}\left( \mathbf{x}\right) \cdot \frac{\partial}{%
\partial\mathbf{x}}V\left( \mathbf{x}\right) dt  \notag
\end{align}
Invoking the FDT, the last term is 
\begin{align}
\int_{0}^{T_{s}}\frac{d\mathbf{x}}{dt}\cdot\mathbf{D}^{-1}\left( \mathbf{x}%
\right) \cdot\mathbf{L}\left( \mathbf{x}\right) \cdot \frac{\partial}{%
\partial\mathbf{x}}V\left( \mathbf{x}\right) dt & =\int _{0}^{T_{s}}\frac{d%
\mathbf{x}}{dt}\cdot\frac{\partial}{\partial\mathbf{x}}V\left( \mathbf{x}%
\right) dt \\
& =\int_{0}^{T_{s}}\frac{d}{dt}V\left( \mathbf{x}\right) dt  \notag \\
& =V\left( \mathbf{x}_{s}\right) -V\left( \mathbf{x}_{1}\right)  \notag
\end{align}
One can therefore write%
\begin{align}
S^{I}\left[ x\right] & =\int_{0}^{T_{s}}\left( \frac {d\mathbf{x}}{dt}\cdot%
\mathbf{D}^{-1}\left( \mathbf{x}\right) \cdot \frac{d\mathbf{x}}{dt}+\left( 
\mathbf{L}\left( \mathbf{x}\right) \cdot \frac{\partial}{\partial\mathbf{x}}%
V\left( \mathbf{x}\right) \right) \cdot\mathbf{D}^{-1}\left( \mathbf{x}%
\right) \cdot\left( \mathbf{L}\left( \mathbf{x}\right) \cdot\frac{\partial}{%
\partial\mathbf{x}}V\left( \mathbf{x}\right) \right) \right) dt \\
& +2V\left( \mathbf{x}_{s}\right) -V\left( \mathbf{x}_{1}\right)  \notag \\
& =\int_{0}^{T_{s}}\left( \frac{d\mathbf{x}}{dt}+\mathbf{L}\left( \mathbf{x}%
\right) \cdot\frac{\partial}{\partial\mathbf{x}}V\left( \mathbf{x}\right)
\right) \cdot\mathbf{D}^{-1}\left( \mathbf{x}\right) \cdot\left( \frac{d%
\mathbf{x}}{dt}+\mathbf{L}\left( \mathbf{x}\right) \cdot\frac{\partial}{%
\partial\mathbf{x}}V\left( \mathbf{x}\right) \right) dt+4\left( V\left( 
\mathbf{x}_{s}\right) -V\left( \mathbf{x}_{1}\right) \right)  \notag
\end{align}
Notice the change in sign of the gradient term. Reversing the sign of the
integration variable and introducing $\mathbf{y}\left( t\right) =\mathbf{x}%
\left( -t\right) $, so that $\mathbf{y}\left( 0\right) =\mathbf{x}_{0}$ and $%
\mathbf{y}\left( -T_{s}\right) =\mathbf{x}_{1}$the action can be written as%
\begin{equation}
S^{I}\left[ x\right] =\int_{-T_{s}}^{0}\left( \frac{d\mathbf{y}}{dt^{\prime}}%
-\mathbf{L}\left( \mathbf{y}\right) \cdot\frac{\partial }{\partial\mathbf{y}}%
V\left( \mathbf{y}\right) \right) \cdot\mathbf{D}^{-1}\left( \mathbf{y}%
\right) \cdot\left( \frac{d\mathbf{y}}{dt^{\prime}}-\mathbf{L}\left( \mathbf{%
y}\right) \cdot\frac{\partial}{\partial \mathbf{y}}V\left( \mathbf{y}\right)
\right) dt^{\prime}+4\left( V\left( \mathbf{x}_{s}\right) -V\left( \mathbf{x}%
_{1}\right) \right)
\end{equation}
The integral is now the action for a path going from $\mathbf{x}_{1}$ at $%
t^{\prime}=-T_{s}$ to $\mathbf{x}_{0}$ at $t^{\prime}=0$ so that the
deterministic path can again be used to set the integral to zero and this
will also determine $T_{s}$. One therefore gets that 
\begin{equation}
S\geq4\left( V\left( \mathbf{x}_{s}\right) -V\left( \mathbf{x}_{1}\right)
\right)
\end{equation}
with equality if the system follows the deterministic paths connecting $%
\mathbf{x}_{s}$ to the end points $\mathbf{x}_{1}$ and $\mathbf{x}_{2}$.
Finally, the action is minimized by choosing $\mathbf{x}_{s}$ to be the
minimal value on the separatrix which is just the critical point.

Note that recrossing the separatrix will involve a deviation from the
deterministic path on both sides of the separatrix and so will not minimize
the action. For this reason, one need only consider a single crossing of the
separatrix.

This serves to establish the claim that for the dynamics given by Eq.(\ref{D}%
) with the fluctuation-dissipation relation, Eq.(\ref{FDT}), the MLP\
crosses the separatrix at the critical point and follows the deterministic
path away from the critical point. Note the key role played by the FDT in
this process: the same result will not necessarily hold in the general case.

\section{Exact action with order parameters\label{AppExact}}

Expanding the Lagrangian in Eq.(\ref{Lagrangian}) gives 
\begin{eqnarray}
\mathcal{L} &=&\frac{1}{2}\int_{0}^{\infty }\frac{1}{r^{2}\rho \left(
r\right) }\left( \frac{\partial m\left( r\right) }{\partial t}\right) ^{2}dr
\\
&&-D\int_{0}^{\infty }\left( \frac{\partial m\left( r\right) }{\partial t}%
\frac{\partial }{\partial r}\frac{1}{r^2}\frac{\delta F\left[ \rho \right] }{\delta \rho
\left( r\right) }\right) dr  \notag \\
&&+\frac{1}{2}D^{2}\int_{0}^{\infty }r^{2}\rho \left( r\right) \left( \frac{%
\partial }{\partial r}\frac{1}{r^2}\frac{\delta F\left[ \rho \right] }{\delta \rho \left(
r\right) }\right) ^{2}dr  \notag
\end{eqnarray}%
If the density is parametrized as $\rho (r;t)=\rho (r;\mathbf{x}(t))$, then
the first term becomes%
\begin{eqnarray}
\frac{1}{2}\int_{0}^{\infty }\frac{1}{r^{2}\rho \left( r\right) }\left( 
\frac{\partial m\left( r\right) }{\partial t}\right) ^{2}dr &=&\left[ \frac{1%
}{2}\int_{0}^{\infty }\frac{1}{r^{2}\rho \left( r;x\right) }\frac{\partial
m\left( r\right) }{\partial x_{a}}\frac{\partial m\left( r\right) }{\partial
x_{b}}dr\right] \frac{dx_{a}}{dt}\frac{dx_{b}}{dt} \\
&=&2\pi g_{ab}\left( \mathbf{x}\right) \frac{dx_{a}}{dt}\frac{dx_{b}}{%
dt}  \notag
\end{eqnarray}%
and the second is%
\begin{eqnarray}
&&-D\int_{0}^{\infty }\left( \frac{\partial m\left( r\right) }{\partial t}%
\frac{\partial }{\partial r}\frac{1}{r^2}\frac{\delta F\left[ \rho \right] }{\delta \rho
\left( r\right) }\right) dr \\
&=&-D\int_{0}^{\infty }\frac{\partial }{\partial
r}\left( \frac{\partial m\left( r\right) }{\partial t}4\pi\left. \frac{\delta F%
\left[ \rho \right] }{\delta \rho \left( \mathbf{r}\right) }\right\vert _{\rho
\left( r\right) }\right) dr+D\int_{0}^{\infty }\left( \frac{\partial }{%
\partial r}\frac{\partial m\left( r\right) }{\partial t}\right) 4\pi\left. \frac{%
\delta F\left[ \rho \right] }{\delta \rho \left( \mathbf{r}\right) }\right\vert
_{\rho \left( r\right) }dr \notag \\
&=&-4\pi D\frac{d x_{a}}{d t}\lim_{r\rightarrow \infty }\left( 
\frac{\partial m\left( r\right) }{\partial x_{a}}\left. \frac{\delta F\left[
\rho \right] }{\delta \rho \left( \mathbf{r}\right) }\right\vert _{\rho \left(
r\right) }\right) +4\pi D\frac{d x_{a}}{d t}\int_{0}^{\infty }4\pi
r^{2}\frac{\partial \rho \left( r\right) }{\partial x_{a}}\left. \frac{%
\delta F\left[ \rho \right] }{\delta \rho \left( \mathbf{r}\right) }\right\vert
_{\rho \left( r\right) }dr  \notag \\
&=&-4\pi D\frac{d x_{a}}{d t}\frac{\partial N}{\partial x_{a}}\mu +4\pi D%
\frac{d x_{a}}{d t}\frac{\partial F\left[ \rho \right] }{%
\partial x_{a}}  \notag
\end{eqnarray}%
To evaluate this, note that $\lim_{r\rightarrow \infty }m(r)=N$, the total
number of particles in the system. Assuming that the boundary condition is
that the density assume the bulk value, $\rho _{\infty }$, far from the
interface gives%
\begin{equation}
\lim_{r\rightarrow \infty }\left( \frac{\partial m\left( r\right) }{\partial
x_{a}}\left. \frac{\delta F\left[ \rho \right] }{\delta \rho \left( \mathbf{r%
}\right) }\right\vert _{\rho \left( r\right) }\right) =\frac{\partial N}{%
\partial x_{a}}\left. \frac{\delta F\left[ \rho \right] }{\delta \rho \left( 
\mathbf{r}\right) }\right\vert _{\rho _{\infty }}\equiv\frac{\partial N}{\partial
x_{a}}\mu
\end{equation}%
where the last equality defines the chemical potential. Combining, the final
result is 
\begin{eqnarray}
-D\int_{0}^{\infty }\left( \frac{\partial m\left( r\right) }{\partial t}%
\frac{\partial }{\partial r}\frac{\delta F\left[ \rho \right] }{\delta \rho
\left( r\right) }\right) dr &=&4\pi D\frac{d x_{a}}{d t}\frac{%
\partial \Omega \left[ \rho \right] }{\partial x_{a}}  
\end{eqnarray}%
where $\Omega \left[ \rho \right] =F\left[ \rho \right] -\mu N$ is the grand
potential. Thus, the Lagrangian is%
\begin{equation}
\frac{1}{4\pi}\mathcal{L}=\frac{1}{2}g_{ab}\left( \mathbf{x}\right) \frac{dx_{a}}{dt}\frac{%
dx_{b}}{dt}+D\frac{d x_{a}}{d t}\frac{\partial \beta \Omega %
\left[ \rho \right] }{\partial x_{a}}+V\left( \mathbf{x}\right)
\end{equation}%
with%
\begin{equation}
V\left( \mathbf{x}\right) =\frac{1}{2}D^{2}\int_{0}^{\infty }r^{2}\rho
\left( r\right) \left( \frac{\partial }{\partial r}\frac{1}{r^2}\frac{\delta F\left[ \rho %
\right] }{\delta \rho \left( r\right) }\right) ^{2}dr
\end{equation}

\section{Eq.(\protect\ref{complete}) as a completeness relation\label%
{AppComplete}}

Assume that $\left\{ p_{i}\left( r\right) \right\} _{i=1}^{\infty }$ is a
complete set of basis functions so that one can write%
\begin{eqnarray}
m\left( r\right) &=&\sum_{i=1}^{\infty }x_{i}p_{i}\left( r\right) \\
x_{i} &=&\int_{0}^{\infty }m\left( r\right) q_{i}\left( r\right) dr  \notag
\end{eqnarray}%
where $\left\{ q_{i}\left( r\right) \right\} _{i=1}^{\infty }$ is the
bi-orthogonal set satisfying%
\begin{equation}
\int_{0}^{\infty }p_{i}\left( r\right) q_{j}\left( r\right) dr=\delta _{ij}
\end{equation}%
Then, since $m(r)$is arbitrary, one has the completeness relation%
\begin{equation}
\sum_{i}p_{i}\left( r\right) q_{i}\left( r^{\prime }\right) =\delta \left(
r-r^{\prime }\right)
\end{equation}%
First note that the metric is%
\begin{equation}
g_{ij}=\int_{0}^{\infty }r^{-2}\rho ^{-1}\left( r\right) \frac{\partial
m\left( r\right) }{\partial x_{i}}\frac{\partial m\left( r\right) }{\partial
x_{j}}dr=\int_{0}^{\infty }r^{-2}\rho ^{-1}\left( r\right) p_{i}\left(
r\right) p_{j}\left( r\right) dr
\end{equation}%
The inverse metric is then%
\begin{eqnarray}
g_{jl}^{-1} &=&\int_{0}^{\infty }r^{\prime 2}\rho \left( r^{\prime }\right) 
\frac{\delta x_{j}}{\delta m\left( r^{\prime }\right) }\frac{\delta x_{l}}{%
\delta m\left( r^{\prime }\right) }dr^{\prime } \\
&=&\int_{0}^{\infty }r^{\prime 2}\rho \left( r^{\prime }\right) q_{j}\left(
r^{\prime }\right) q_{l}\left( r^{\prime }\right) dr^{\prime }  \notag
\end{eqnarray}%
as can be verified by direct evaluation of 
\begin{eqnarray}
&&\sum_{j}g_{ij}\int_{0}^{\infty }r^{\prime 2}\rho \left( r^{\prime }\right)
q_{j}\left( r^{\prime }\right) q_{l}\left( r^{\prime }\right) dr^{\prime } \\
&=&\sum_{j}\int_{0}^{\infty }dr\int_{0}^{\infty }dr^{\prime }\;\left(
r^{-2}\rho ^{-1}\left( r\right) p_{i}\left( r\right) p_{j}\left( r\right)
\right) \left( r^{\prime 2}\rho \left( r^{\prime }\right) q_{j}\left(
r^{\prime }\right) q_{l}\left( r^{\prime }\right) \right) \notag \\
&=&\int_{0}^{\infty }dr\int_{0}^{\infty }dr^{\prime }\;\left( r^{-2}\rho
^{-1}\left( r\right) p_{i}\left( r\right) \right) \delta \left( r-r^{\prime
}\right) \left( r^{\prime 2}\rho \left( r^{\prime }\right) q_{l}\left(
r^{\prime }\right) \right)  \notag \\
&=&\int_{0}^{\infty }dr\;p_{i}\left( r\right) q_{l}\left( r^{\prime }\right)
\notag \\
&=&\delta _{il}  \notag
\end{eqnarray}%
where the second line follows from the completeness relation and the last
line by bi-orthogonality. Then, one easily verifies that 
\begin{eqnarray}
\sum_{ij}\frac{\partial m\left( r\right) }{\partial x_{i}}g_{ij}^{-1}\frac{%
\partial m\left( r^{\prime }\right) }{\partial x_{j}} &=&\sum_{ij}p_{i}%
\left( r\right) \left( \int_{0}^{\infty }r^{\prime \prime 2}\rho \left(
r^{\prime \prime }\right) q_{i}\left( r^{\prime \prime }\right) q_{j}\left(
r^{\prime \prime }\right) dr^{\prime \prime }\right) p_{j}\left( r^{\prime
}\right) \\
&=&\int_{0}^{\infty }r^{\prime \prime 2}\rho \left( r^{\prime \prime
}\right) \delta \left( r-r^{\prime \prime }\right) \delta \left( r^{\prime
}-r^{\prime \prime }\right) dr^{\prime \prime }  \notag \\
&=&r^{2}\rho \left( r\right) \delta \left( r-r^{\prime }\right)  \notag
\end{eqnarray}%
as claimed.

\section{Alternative derivation of order parameter equations}

\label{AltDerivation}

One way to relate a general functional form, $\widetilde{\rho }\left( r;%
\mathbf{x}\left( t\right) \right) $, to the exact density, $\rho \left(
r;t\right) $, is by defining a distance in density space and then minimizing
the difference between the two functions. It is shown in the main text that
the dynamics imposes a unique distance functional which is expressed in
terms of the corresponding mass functionals as%
\begin{equation}
s^{2}=\int_{0}^{\infty }\left( \widetilde{m}^{-1}\left( u;\mathbf{x}\left(
t\right) \right) -m^{-1}\left( u;t\right) \right) ^{2}du
\end{equation}%
Determining the fitting parameters at each moment in time by minimizing the
difference between the actual mass distribution and the approximating
distribution gives%
\begin{eqnarray}
0 &=&\frac{\partial }{\partial x_{i}\left( t\right) }\int_{0}^{\infty
}\left( \widetilde{m}^{-1}\left( u;\mathbf{x}\left( t\right) \right)
-m^{-1}\left( u;t\right) \right) ^{2}du  \label{app1} \\
&=&\int_{0}^{\infty }\left( \widetilde{m}^{-1}\left( u;\mathbf{x}\left(
t\right) \right) -m^{-1}\left( u;t\right) \right) \frac{\partial \widetilde{m%
}^{-1}\left( u;\mathbf{x}\left( t\right) \right) }{\partial x_{i}\left(
t\right) }du  \notag \\
&=&\int_{0}^{\infty }\left( \widetilde{m}^{-1}\left( m\left( r;t\right) ;%
\mathbf{x}\left( t\right) \right) -r\right) \frac{\partial \widetilde{m}%
^{-1}\left( u;\mathbf{x}\left( t\right) \right) }{\partial x_{i}\left(
t\right) }\frac{\partial m\left( r;t\right) }{\partial r}dr  \notag
\end{eqnarray}%
In the following, in order to simplify the notation, the dependence on time
will not be explicitly indicated.

Noting that if%
\begin{equation}
y\left( m;\mathbf{x}\right) =\widetilde{m}^{-1}\left( m;\mathbf{x}\right)
\end{equation}%
then%
\begin{equation}
m=\widetilde{m}\left( y\left( m;\mathbf{x}\right) ;\mathbf{x}\right)
\end{equation}%
so%
\begin{equation}
0=\widetilde{m}_{r}\left( y\left( m,\mathbf{x}\right) ;\mathbf{x}\right)
y_{i}\left( m;\mathbf{x}\right) +\widetilde{m}_{i}\left( y\left( m;\mathbf{x}%
\right) ;\mathbf{x}\right)
\end{equation}%
where the short-hand notation%
\begin{eqnarray}
\widetilde{m}_{r}\left( y;\mathbf{x}\right) &\equiv &\left. \frac{\partial 
\widetilde{m}\left( r;\mathbf{x}\right) }{\partial r}\right\vert _{r=y} \\
\widetilde{m}_{i}\left( y;\mathbf{x}\right) &\equiv &\frac{\partial 
\widetilde{m}\left( y;\mathbf{x}\right) }{\partial x_{i}}  \notag
\end{eqnarray}%
is used. Thus, Eq.(\ref{app1}) becomes 
\begin{equation}
0=\int_{0}^{\infty }\left( \widetilde{m}^{-1}\left( m\left( r\right) ;%
\mathbf{x}\right) -r\right) \frac{\widetilde{m}_{i}\left( \widetilde{m}%
^{-1}\left( m\left( r\right) ;\mathbf{x}\right) ;\mathbf{x}\right) }{%
\widetilde{m}_{r}\left( \widetilde{m}^{-1}\left( m\left( r\right) ;\mathbf{x}%
\right) ;\mathbf{x}\right) }\frac{\partial m\left( r\right) }{\partial r}dr
\end{equation}%
Another derivative gives%
\begin{eqnarray}
0 &=&-\frac{dx_{j}}{dt}\int_{0}^{\infty }\frac{\widetilde{m}_{j}\left( 
\widetilde{m}^{-1}\left( m\left( r\right) ;\mathbf{x}\right) ;\mathbf{x}%
\right) }{\widetilde{m}_{r}\left( \widetilde{m}^{-1}\left( m\left( r\right) ;%
\mathbf{x}\right) ;\mathbf{x}\right) }\frac{\widetilde{m}_{i}\left( 
\widetilde{m}^{-1}\left( m\left( r\right) ;\mathbf{x}\right) ;\mathbf{x}%
\right) }{\widetilde{m}_{r}\left( \widetilde{m}^{-1}\left( m\left( r\right) ;%
\mathbf{x}\right) ;\mathbf{x}\right) }\frac{\partial m\left( r\right) }{%
\partial r}dr \\
&&+\int_{0}^{\infty }\widetilde{m}_{r}^{-1}\left( m\left( r\right) ;\mathbf{x%
}\right) \frac{dm\left( r\right) }{dt}\frac{\widetilde{m}_{i}\left( 
\widetilde{m}^{-1}\left( m\left( r\right) ;\mathbf{x}\right) ;\mathbf{x}%
\right) }{\widetilde{m}_{r}\left( \widetilde{m}^{-1}\left( m\left( r\right) ;%
\mathbf{x}\right) ;\mathbf{x}\right) }\frac{\partial m\left( r\right) }{%
\partial r}dr  \notag \\
&&+\frac{dx_{j}}{dt}\int_{0}^{\infty }\left( \widetilde{m}^{-1}\left(
m\left( r\right) ;\mathbf{x}\right) -r\right) \frac{d}{dt}\left[ \frac{%
\widetilde{m}_{i}\left( \widetilde{m}^{-1}\left( m\left( r\right) ;\mathbf{x}%
\right) ;\mathbf{x}\right) }{\widetilde{m}_{r}\left( \widetilde{m}%
^{-1}\left( m\left( r\right) ,\mathbf{x}\right) ;\mathbf{x}\right) }\frac{%
\partial m\left( r\right) }{\partial r}\right] dr  \notag
\end{eqnarray}%
or%
\begin{eqnarray}
0 &=&-\frac{dx_{j}}{dt}\int_{0}^{\infty }\frac{\widetilde{m}_{j}\left( 
\widetilde{m}^{-1}\left( m\left( r\right) ;\mathbf{x}\right) ;\mathbf{x}%
\right) }{\widetilde{m}_{r}\left( \widetilde{m}^{-1}\left( m\left( r\right) ;%
\mathbf{x}\right) ;\mathbf{x}\right) }\frac{\widetilde{m}_{i}\left( 
\widetilde{m}^{-1}\left( m\left( r\right) ;\mathbf{x}\right) ;\mathbf{x}%
\right) }{\widetilde{m}_{r}\left( \widetilde{m}^{-1}\left( m\left( r\right) ;%
\mathbf{x}\right) ;\mathbf{x}\right) }\frac{\partial m\left( r\right) }{%
\partial r}dr \\
&&+\int_{0}^{\infty }\frac{1}{\widetilde{m}_{r}\left( \widetilde{m}%
^{-1}\left( m\left( r\right) ;\mathbf{x}\right) ;\mathbf{x}\right) }\frac{%
dm\left( r\right) }{dt}\frac{\widetilde{m}_{i}\left( \widetilde{m}%
^{-1}\left( m\left( r\right) ;\mathbf{x}\right) ;\mathbf{x}\right) }{%
\widetilde{m}_{r}\left( \widetilde{m}^{-1}\left( m\left( r\right) ;\mathbf{x}%
\right) ;\mathbf{x}\right) }\frac{\partial m\left( r\right) }{\partial r}dr 
\notag \\
&&+\frac{dx_{j}}{dt}\int_{0}^{\infty }\left( \widetilde{m}^{-1}\left(
m\left( r\right) ;\mathbf{x}\right) -r\right) \frac{d}{dt}\left[ \frac{%
\widetilde{m}_{i}\left( \widetilde{m}^{-1}\left( m\left( r\right) ;\mathbf{x}%
\right) ;\mathbf{x}\right) }{\widetilde{m}_{r}\left( \widetilde{m}%
^{-1}\left( m\left( r\right) ,\mathbf{x}\right) ;\mathbf{x}\right) }\frac{%
\partial m\left( r\right) }{\partial r}\right] dr  \notag
\end{eqnarray}%
Expanding in the difference $m\left( r\right) -\widetilde{m}\left( r;\mathbf{%
x}\right) $ gives 
\begin{equation}
g_{ij}\left( \mathbf{x}\left( t\right) \right) \frac{dx_{i}\left( t\right) }{%
dt}=\int_{0}^{\infty }\frac{\partial \widetilde{m}\left( r;\mathbf{x}\left(
t\right) \right) }{\partial x_{i}\left( t\right) }\left( \frac{\partial
m\left( r;t\right) }{\partial r}\right) ^{-1}\frac{dm\left( r;t\right) }{dt}%
dr+O\left( m\left( r;t\right) -\widetilde{m}\left( r;\mathbf{x}\left(
t\right) \right) \right)
\end{equation}%
which gives the result, Eq.(\ref{final}).

\section{The spurious drift\label{Noise}}

The stochastic model is 
\begin{equation}
\frac{dx_{i}}{dt}=-Dg_{ij}^{-1}\left( \mathbf{x}\right) \frac{\partial \beta
\Omega }{\partial x_{j}}-\epsilon g_{ij}^{-1}\left( \mathbf{x}\right) \int_{0}^{\infty} 
\sqrt{\frac{2D}{4\pi r^{2}\rho\left( r;\mathbf{x}\right) }}\frac{\partial
m\left( r;\mathbf{x}\right) }{\partial x_{j}}\xi \left( r;t\right) dr.
\end{equation}%
which must be understood in the Stratonovich interpretation. In order to
replace the noise term by a simpler form, this must first be written as an
Ito SDE. Then, the change of the noise can be made and the result
transformed back to the Stratonovich interpretation. The first step is
accomplished using the standard transformation rule\cite{Gardiner} giving the equivalent Ito form of the SDE%
\begin{eqnarray}
\frac{dx_{i}}{dt} &=&-Dg_{ij}^{-1}\left( \mathbf{x}\right) \frac{\partial
\beta \Omega }{\partial x_{j}} \label{F2} \\
&&+\frac{1}{2}\epsilon ^{2}\int_{0}^{\infty} \left( \frac{\partial }{\partial x_{l}}%
g_{ij}^{-1}\left( \mathbf{x}\right) \sqrt{\frac{2D}{4\pi r^{2}\rho\left( r;%
\mathbf{x}\right) }}\frac{\partial m\left( r;\mathbf{x}\right) }{\partial
x_{j}}\right) \left( g_{lk}^{-1}\left( \mathbf{x}\right) \sqrt{\frac{2D}{%
4\pi r^{2}\rho\left( r;\mathbf{x}\right) }}\frac{\partial m\left( r;\mathbf{x}%
\right) }{\partial x_{k}}\right) dr  \notag \\
&&-\epsilon g_{ij}^{-1}\left( \mathbf{x}\right) \int_{0}^{\infty} \sqrt{\frac{2D}{4\pi
r^{2}\rho\left( r;\mathbf{x}\right) }}\frac{\partial m\left( r;\mathbf{x}%
\right) }{\partial x_{j}}\xi \left( r;t\right) dr.  \notag
\end{eqnarray}%
The autocorrelation of the noise is
\begin{eqnarray}
D_{il}&\equiv& \epsilon^2g_{ij}^{-1}g_{lm}^{-1}\int_{0}^{\infty}\frac{2D}{4\pi r^2 \rho\left( r;\mathbf{x}\right)} \frac{\partial m\left( r;\mathbf{x}\right)}{\partial x_j}\frac{\partial m\left( r;\mathbf{x}\right)}{\partial x_m}dr \\
&=& 2D\epsilon^2 g_{il}^{-1}
\end{eqnarray}
so that the noise term in Eq.(\ref{F2}) can be replace by the equivalent form $\epsilon\sqrt{2D}q^{-1}_{ij}(\mathbf{x})\xi_j(t)$ with $q_{il}(\mathbf{x})q_{jl}(\mathbf{x})=g_{ij}(\mathbf{x})$ and $\left<\xi_i(t)\xi_j(t')\right>=\delta_{ij}\delta(t-t')$.

Up to a factor of $D$, the second term on the right is%
\begin{eqnarray}
A_{i}^{\left( I\right) } &=&\frac{\partial g_{ij}^{-1}\left( \mathbf{x}%
\right) }{\partial x_{j}} \\
&&+g_{ij}^{-1}\left( \mathbf{x}\right) g_{lk}^{-1}\left( \mathbf{x}\right)
\int_{0}^{\infty} \frac{1}{4\pi r^{2}\rho\left( r;\mathbf{x}\right) }\left( \frac{\partial
^{2}m\left( r;\mathbf{x}\right) }{\partial x_{l}\partial x_{i}}-\frac{1}{2}%
\frac{1}{\rho\left( r;\mathbf{x}\right) }\frac{\partial \rho\left( r;\mathbf{x}%
\right) }{\partial x_{l}}\frac{\partial m\left( r;\mathbf{x}\right) }{%
\partial x_{i}}\right) \frac{\partial m\left( r;\mathbf{x}\right) }{\partial
x_{k}}dr.  \notag
\end{eqnarray}%
Using the symmetry of $g_{ij}^{-1}$ one has that%
\begin{eqnarray}
&&g_{ij}^{-1}\left( \mathbf{x}\right) g_{lk}^{-1}\left( \mathbf{x}\right)
\int_{0}^{\infty} \frac{1}{4\pi r^{2}\rho\left( r;\mathbf{x}\right) }\frac{\partial
^{2}m\left( r;\mathbf{x}\right) }{\partial x_{l}\partial x_{i}}\frac{%
\partial m\left( r;\mathbf{x}\right) }{\partial x_{k}}dr \\
&=&\frac{1}{2}g_{ij}^{-1}\left( \mathbf{x}\right) g_{lk}^{-1}\left( \mathbf{x%
}\right) \int_{0}^{\infty} \frac{1}{4\pi r^{2}\rho\left( r;\mathbf{x}\right) }\left( \frac{%
\partial ^{2}m\left( r;\mathbf{x}\right) }{\partial x_{l}\partial x_{i}}%
\frac{\partial m\left( r;\mathbf{x}\right) }{\partial x_{k}}+\frac{\partial
^{2}m\left( r;\mathbf{x}\right) }{\partial x_{k}\partial x_{i}}\frac{%
\partial m\left( r;\mathbf{x}\right) }{\partial x_{l}}\right) dr  \notag \\
&=&\frac{1}{2}g_{ij}^{-1}\left( \mathbf{x}\right) g_{lk}^{-1}\left( \mathbf{x%
}\right) \int_{0}^{\infty} \frac{1}{4\pi r^{2}\rho\left( r;\mathbf{x}\right) }\frac{\partial 
}{\partial x_{i}}\left( \frac{\partial m\left( r;\mathbf{x}\right) }{%
\partial x_{l}}\frac{\partial m\left( r;\mathbf{x}\right) }{\partial x_{k}}%
\right) dr  \notag \\
&=&\frac{1}{2}g_{ij}^{-1}\left( \mathbf{x}\right) g_{lk}^{-1}\left( \mathbf{x%
}\right) \left[ \frac{\partial g_{lk}\left( \mathbf{x}\right) }{\partial
x_{i}}+\int_{0}^{\infty} \frac{1}{4\pi r^{2}\rho^{2}\left( r;\mathbf{x}\right) }\frac{%
\partial \rho\left( r;\mathbf{x}\right) }{\partial x_{i}}\frac{\partial m\left(
r;\mathbf{x}\right) }{\partial x_{l}}\frac{\partial m\left( r;\mathbf{x}%
\right) }{\partial x_{k}}dr\right]  \notag
\end{eqnarray}%
so the extra term becomes%
\begin{eqnarray}
A_{i}^{\left( I\right) } &=&\frac{\partial g_{ij}^{-1}\left( \mathbf{x}%
\right) }{\partial x_{j}}+\frac{1}{2}g_{ij}^{-1}\left( \mathbf{x}\right)
g_{lk}^{-1}\left( \mathbf{x}\right) \frac{\partial g_{lk}\left( \mathbf{x}%
\right) }{\partial x_{i}} \\
&&+\frac{1}{2}g_{ij}^{-1}\left( \mathbf{x}\right) g_{lk}^{-1}\left( \mathbf{x%
}\right) \int_{0}^{\infty} \frac{1}{4\pi r^{2}f^{2}\left( r;\mathbf{x}\right) }\left( 
\frac{\partial \rho\left( r;\mathbf{x}\right) }{\partial x_{i}}\frac{\partial
m\left( r;\mathbf{x}\right) }{\partial x_{l}}-\frac{\partial \rho\left( r;%
\mathbf{x}\right) }{\partial x_{l}}\frac{\partial m\left( r;\mathbf{x}%
\right) }{\partial x_{i}}\right) \frac{\partial m\left( r;\mathbf{x}\right) 
}{\partial x_{k}}dr  \notag \\
&=&\frac{\partial g_{ij}^{-1}\left( \mathbf{x}\right) }{\partial x_{j}}+%
\frac{1}{2}g_{ij}^{-1}\left( \mathbf{x}\right) g_{lk}^{-1}\left( \mathbf{x}%
\right) \frac{\partial g_{lk}\left( \mathbf{x}\right) }{\partial x_{i}} 
\notag \\
&&+\frac{1}{2}\left( g_{ij}^{-1}\left( \mathbf{x}\right) g_{lk}^{-1}\left( 
\mathbf{x}\right) -g_{lj}^{-1}\left( \mathbf{x}\right) g_{ik}^{-1}\left( 
\mathbf{x}\right) \right) \int_{0}^{\infty} \frac{1}{4\pi r^{2}\rho^{2}\left( r;\mathbf{x}%
\right) }\frac{\partial \rho\left( r;\mathbf{x}\right) }{\partial x_{i}}\frac{%
\partial m\left( r;\mathbf{x}\right) }{\partial x_{l}}\frac{\partial m\left(
r;\mathbf{x}\right) }{\partial x_{k}}dr.  \notag
\end{eqnarray}%
Combining these results gives the equivalent Ito form of the SDE with simplified noise term
\begin{equation}
\frac{dx_{i}}{dt}=-2Dg_{ij}^{-1}\left( \mathbf{x}\right) \frac{\partial
\beta \Omega }{\partial x_{j}}-D\epsilon ^{2}A_{i}^{\left( I\right)
}-\epsilon \sqrt{2D}q_{lj}^{-1}\left( \mathbf{x}\right) \xi _{l}\left(
t\right).
\end{equation}%
To get the Stratonovich form, one must transform back giving the same
equation but with spurious drift%
\begin{equation}
A_{i}^{\left( S\right) }=A_{i}^{\left( I\right) }-q_{kj}^{-1}\frac{\partial
q_{ij}^{-1}}{\partial x_{k}}.
\end{equation}

\end{document}